\documentclass[10pt,conference]{IEEEtran}

\IEEEoverridecommandlockouts

\usepackage{cite}
\usepackage{amsmath,amssymb,amsfonts}
\usepackage{algorithmicx}
\usepackage{algpseudocode}
\usepackage{graphicx}
\usepackage{textcomp}
\usepackage[shortlabels]{enumitem}
\def\BibTeX{{\rm B\kern-.05em{\sc i\kern-.025em b}\kern-.08em
    T\kern-.1667em\lower.7ex\hbox{E}\kern-.125emX}}
    
\usepackage{hyperref}
\usepackage[ruled, linesnumbered, vlined, commentsnumbered]{algorithm2e}
\usepackage{amsmath,blkarray,booktabs}
\usepackage{xcolor,colortbl}
\usepackage{multirow}
\usepackage{subcaption} 
\usepackage{tcolorbox}
\usepackage{csquotes}
\usepackage{tikz,pgfplots}
\usepackage{standalone}
\usepackage{pgfplots}
\usepackage{graphicx}
\usepackage{balance}
\usepackage{siunitx}
\usepackage{pifont}
\usepackage{adjustbox}
\usepackage{makecell}
\usepackage{listings}

\usepgfplotslibrary{fillbetween}
\usetikzlibrary{fit,shapes.arrows}

\pgfplotsset{
 /pgfplots/ybar legend/.style={
        /pgfplots/legend image code/.code={
            \draw [##1,/tikz/.cd] (0cm,-0.3em) rectangle (0.2em,0.4em);
        },
    },
}

\lstdefinelanguage{json}{
  basicstyle=\ttfamily\footnotesize,
  numbers=left,
  numberstyle=\tiny\color{gray},
  stepnumber=1,
  numbersep=8pt,
  showstringspaces=false,
  breaklines=true,
  frame=single,
  backgroundcolor=\color{gray!5},
  literate=
   *{0}{{{\textcolor{blue}{0}}}}{1}
    {1}{{{\textcolor{blue}{1}}}}{1}
    {2}{{{\textcolor{blue}{2}}}}{1}
    {3}{{{\textcolor{blue}{3}}}}{1}
    {4}{{{\textcolor{blue}{4}}}}{1}
    {5}{{{\textcolor{blue}{5}}}}{1}
    {6}{{{\textcolor{blue}{6}}}}{1}
    {7}{{{\textcolor{blue}{7}}}}{1}
    {8}{{{\textcolor{blue}{8}}}}{1}
    {9}{{{\textcolor{blue}{9}}}}{1}
    {:}{{{\textcolor{black}{:}}}}{1}
    {,}{{{\textcolor{black}{,}}}}{1}
    {\{}{{{\textcolor{black}{\{}}}}{1}
    {\}}{{{\textcolor{black}{\}}}}}{1}
    {[}{{{\textcolor{black}{[}}}}{1}
    {]}{{{\textcolor{black}{]}}}}{1},
}

\newcommand{\Model}{\texttt{CoTune}} 
\newcommand{\tuner}{\texttt{CoTune}}

\newcommand{\vect}[1]{\boldsymbol{#1}}

\newcommand{\revisioncolor}{black} 
\newcommand{\revision}[1]{{\color{\revisioncolor}#1}}

\newcommand{\sta}[1]{
\begin{tcolorbox}[colback=gray!20,leftrule=0mm,rightrule=0mm,toprule=0mm,bottomrule=0mm,left=0pt,right=0pt,top=1pt,bottom=1pt]
\em #1
\end{tcolorbox}
}

\newtcolorbox{innerbox}{colback=green!10,boxrule=0.2pt,colframe=black,top=0pt,bottom=0pt,left=1pt,right=1pt}

\newtcolorbox{rbox}{colback=green!10,boxrule=0.2pt,colframe=black,top=1pt,bottom=1pt,left=1pt,right=1pt}

\begin{document}


\title{CoTune: Co-evolutionary Configuration Tuning}

\author{
\IEEEauthorblockN{Gangda Xiong$^{1}$ and Tao Chen$^{2 \ast}$}

\IEEEauthorblockA{$^1$ School of Computer Science and Engineering, University of Electronic Science and Technology of China, China}
\IEEEauthorblockA{$^2$ IDEAS Lab, School of Computer Science, University of Birmingham, United Kingdom}

\IEEEauthorblockA{gangdaxiong0207@gmail.com, t.chen@bham.ac.uk}

\thanks{$^{\ast}$Tao Chen is the corresponding author. Gangda Xiong is also supervised in the IDEAS Lab.}
}

\maketitle

\begin{abstract}

To automatically tune configurations for the best possible system performance (e.g., runtime or throughput), much work has been focused on designing intelligent heuristics in a tuner. However, existing tuner designs have mostly ignored the presence of complex performance requirements (e.g., ``\texttt{the latency shall ideally be 2 seconds}''), but simply assume that better performance is always more preferred. This would not only waste valuable information in a requirement but might also consume extensive resources to tune for a goal with little gain. Yet, prior studies have shown that simply incorporating the requirement as a tuning objective is problematic since the requirement might be too strict, harming convergence; or its highly diverse satisfactions might lead to premature convergence.

In this paper, we propose \tuner, a tool that takes the information of a given target performance requirement into account through co-evolution. \tuner~is unique in the sense that it creates an auxiliary performance requirement to be co-evolved with the configurations, which assists the target performance requirement when it becomes ineffective or even misleading, hence allowing the tuning to be guided by the requirement while being robust to its harm. Experiment results on 162 cases (nine systems and 18 requirements) reveal that \tuner~considerably outperforms existing tuners, ranking as the best for $\approx90\%$ cases (against the $0\%$--$35\%$ for other tuners) with up to $2.9\times$ overall improvements, while doing so under a much better efficiency.

\end{abstract}



\section{Introduction}
\label{sec:introduction}


Modern software systems expose an ever‐growing number of configuration options---ranging from thread/cache sizes to algorithm/architecture choices---offering users unprecedented control over the performance needs, e.g., runtime or throughput~\cite{DBLP:journals/tse/SayaghKAP20,chen2025accuracy,DBLP:journals/tse/Nair0MSA20,DBLP:conf/icse/LiangChen25,DBLP:journals/corr/abs-2501-15392}. Yet this flexibility comes at a steep price: in 2017-2018 alone, severe violations of performance requirements caused by poor configuration have occurred in more than 50\% of the software companies worldwide, costing $\approx400,000$ USD per hour in average~\cite{exp}. As such, configuration tuning is an extremely important task for software quality assurance.

Tuning configuration is not easy, because even a single configuration measurement is expensive, \revision{e.g., it can take up to 166 minutes to measure one configuration on a database system\cite{DBLP:conf/sigsoft/0001L24}}, let alone the exponentially growing configuration space~\cite{DBLP:journals/tse/SayaghKAP20,chen2025accuracy}, \revision{e.g., the recent version of system \textsc{Jump3R} has $6.87\times10^{10}$ configurations.} This renders brute‐force tuning infeasible. To automatically tune the configurations for better performance, the tuner design has focused on smart heuristics, e.g., iterated local search~\cite{bestconfig}, Genetic Algorithm~\cite{DBLP:conf/wcre/Chen22,DBLP:journals/tosem/ChenLBY18,k2vtune,DBLP:conf/icse/YeChen25,DBLP:journals/ase/GerasimouCT18,DBLP:journals/corr/abs-2112-07303,DBLP:conf/wosp/MartensKBR10}, and Bayesian optimization~\cite{DBLP:conf/mascots/JamshidiC16}, together with its variants~\cite{DBLP:journals/tse/Nair0MSA20,SMAC}. 

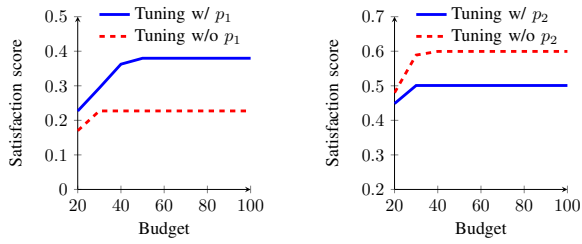
\begin{figure}[!t]
 \centering
  \begin{subfigure}[t]{0.4\columnwidth}
    \centering
\begin{tikzpicture}
    \begin{axis}[
      axis x line  = bottom,
            axis y line  = left  ,
        width=5cm, height=5cm, 
        xlabel={Budget},
        ylabel={Satisfaction score},
        xmin=20, xmax=100,
        xtick={20,40,60,80,100},
        ymin=0, ymax=0.5,
        ytick={0,0.1,0.2,0.3,0.4,0.5},
            legend cell align=left,
          legend columns=1,
        legend style={
          draw=none, fill=none,
          at={(1.0,0.8)}, anchor=south east,
          legend cell align=left
        }
    ]

    \addplot[
        blue,
        mark=none,
        ultra thick
    ] coordinates {
(20,0.22696167)
(30,0.29365986)
(40,0.36252604)
(50,0.37969516)
(60,0.37969516)
(70,0.37969516)
(80,0.37969516)
(90,0.37969516)
(100,0.37969516)
    }; 

    \addplot[
        red,
        mark=none,
        ultra thick,dashed
    ] coordinates {
(20,0.16997001)
(30,0.22721508)
(40,0.22721508)
(50,0.22721508)
(60,0.22721508)
(70,0.22721508)
(80,0.22721508)
(90,0.22721508)
(100,0.22721508)
    }; 

   
\addlegendentry{Tuning w/ $p_1$}
\addlegendentry{Tuning w/o $p_1$}
    \end{axis}
\end{tikzpicture}
  \subcaption{Requirement is helpful}
  \end{subfigure}
 ~\hspace{0.3cm}
  \begin{subfigure}[t]{0.4\columnwidth}
    \centering
\begin{tikzpicture}
    \begin{axis}[
      axis x line  = bottom,
            axis y line  = left  ,
        width=5cm, height=5cm, 
        xlabel={Budget},
        ylabel={Satisfaction score},
        xmin=20, xmax=100,
        xtick={20,40,60,80,100},
        ymin=0.2, ymax=0.7,
        ytick={0.2,0.3,0.4,0.5,0.6,0.7},
            legend cell align=left,
          legend columns=1,
        legend style={
          draw=none, fill=none,
          at={(1.0,0.8)}, anchor=south east,
          legend cell align=left
        }
    ]

    \addplot[
        blue,
        mark=none,
        ultra thick
    ] coordinates {
(20,0.44848337)
(30,0.50085147)
(40,0.50085147)
(50,0.50085147)
(60,0.50085147)
(70,0.50085147)
(80,0.50085147)
(90,0.50085147)
(100,0.50085147)
    }; 

    \addplot[
        red,
        mark=none,
        ultra thick,dashed
    ] coordinates {
(20,0.47983886)
(30,0.58886000)
(40,0.59948863)
(50,0.59948863)
(60,0.59948863)
(70,0.59948863)
(80,0.59948863)
(90,0.59948863)
(100,0.59948863)

    }; 

   \addlegendentry{Tuning w/ $p_2$}
\addlegendentry{Tuning w/o $p_2$}

    \end{axis}
\end{tikzpicture}
    \subcaption{Requirement is harmful}
  \end{subfigure}
  
    \caption{The satisfaction scores of two cases for tuning with and without a requirement on \textsc{SQLite} via a tuner based on Genetic Algorithm. \revision{$p_1$ and $p_2$ denote two different requirements.}}
   \label{fig:pre-exp}
   \vspace{-0.3cm}
 \end{figure}

Despite the rapidly developing tuners, they mostly rely on a simplified assumption: the better the performance, the more preferred, and configuration tuning is all about finding the optimal configuration that achieves the best possible performance~\cite{k2vtune,DBLP:journals/ase/GerasimouCT18,DBLP:journals/corr/abs-2112-07303,DBLP:journals/tse/SayaghKAP20,chen2025accuracy,DBLP:journals/tse/Nair0MSA20}. Indeed, the above is cognitively not incorrect, but the practical scenarios are far more complicated due to the presence of performance requirements~\cite{DBLP:journals/tosem/ChenL23a}: it is not uncommon to see a similar performance need as the one in Figure~\ref{fig:perf-req} documented for a system. There are a couple of reasons that it is insufficient to simply assume the better is constantly more preferred: for example, since the configuration measurement is expensive, it might not always be ideal to find the best possible runtime as the improvement when the configuration is already good enough might be minor but achieving so would consume much more budgets. On the other hand, there could be cases where a runtime worse than a certain point is equally unacceptable, although a better runtime than that point might be preferred. The rich information in a given performance requirements can serve as a valuable source to guide the tuning, better finding what is truly satisfied by stakeholders. Indeed, a study from Chen and Li~\cite{DBLP:journals/tosem/ChenL23a} reveals that fitting the requirement as the objective for configuration tuning generally leads to much higher satisfaction than tuning without (see Figure~\ref{fig:pre-exp}a). Further, ignoring the requirement is risky to spend extra resources for tuning more than necessary.


\begin{figure}[t!]
\centering

\footnotesize

\begin{adjustbox}{width=\columnwidth,center}
\begin{tabular}{lp{8.5cm}}
\toprule
\textbf{Requirement 4.3.1:}&The system should support at least 1000 concurrent users.\\
\midrule
\textbf{Description:}&This statement provides a general sense of reliability when the system is under load. It is important that a substantial number of actors be able to access the system at the same time, since a courseware system is important to the courses that employ it. The times when the system will be under the most stress are likely during midterm and finals weeks. Therefore, it must be able to handle at least 1,000 concurrent users.\\
\bottomrule
\end{tabular}
\end{adjustbox}



\caption{A real-world performance requirement \revision{from the Puget Sound Enhancements System in the \textsc{PURE} dataset~\cite{pure}}.}
\label{fig:perf-req}
\vspace{-0.3cm}
\end{figure}

However, directly using a performance requirement as an objective for a tuner, as what has been done in some existing works~\cite{DBLP:journals/ase/GerasimouCT18,DBLP:journals/corr/abs-2112-07303}, can be problematic. As hinted by Chen and Li~\cite{DBLP:journals/tosem/ChenL23a}, together with our own experiment observations, requirements can sometimes be ineffective or even harmful to the tuning (see Figure~\ref{fig:pre-exp}b): too strict requirements can lead to the loss of search pressure, negatively affecting the convergence of tuning; the highly diverse satisfactions in the requirement might cause stagnation, yielding premature convergence. All those are inevitable since at the requirement elicitation stage, it is difficult to know the feasibility of expectations nor how the performance requirement impacts the tuning.





To address the above gap, in this paper, we present \tuner, the first tuner that dynamically ``co-evolves'' a given performance requirement and configurations for better satisfaction. Unlike others, \tuner~creates an \textbf{\textit{auxiliary}} performance requirement, together with the given \textbf{\textit{target}} one, to guide the tuning: the former is co-evolved with the configurations and assists the latter when it becomes useless or harmful. In this way, we neither statically rely on the fixed target performance requirement nor completely ignore it, but take advantage of the rich information therein while catering for its potential limits and harms on-the-fly during tuning. Our contributions are:

\begin{itemize}
    \item We represent performance requirements in a quantifiable manner via fuzzy logic~\cite{zadeh1988fuzzy} that fits with the tuning context, enabling intuitive specification by the developers.
    \item Through cooperative co-evolution~\cite{DBLP:journals/tec/MaLZTLXZ19}, we co-evolve requirements and configurations based on the tuning status.
    \item We exploit differential entropy to measure the discriminative power of the requirement on configurations---the key to mitigating the loss of pressure and stagnation---paired with the corresponding co-evolution mechanisms.
    \item We evaluate \tuner~against several state-of-the-art tuners, with and without guidance of performance requirement, under nine systems and 18 possible target performance requirements, leading to 162 cases.
\end{itemize}

The results demonstrate that \tuner~is ranked the best for $\approx 90\%$ cases, which is significantly better than the overall best counterpart, with up to $2.9\times$ overall satisfaction improvement and a much superior efficiency. The Code/data can be found at: \textcolor{blue}{\texttt{\url{https://github.com/ideas-labo/CoTune}}}.



The paper is organized as follows:
Section~\ref{sec:background} introduces preliminaries/motivations.  Section~\ref{sec:framework} explains \tuner.  Section~\ref{sec:experiment_setup} describes our experimental setup. Section~\ref{sec:results} analyzes the results followed by a discussion in Section~\ref{sec:discussion}. Section~\ref{sec:related_work} reviews related work and Section~\ref{sec:conclusion} concludes the paper.



\section{Preliminaries}
\label{sec:background}



\subsection{Configuration Tuning}

Software configuration tuning seeks to automatically find a configuration that optimizes a performance metric of the system (e.g., runtime, throughput, energy). Formally, the goal is to find (assuming minimizing performance metric):
\begin{equation}
\label{eq:singleobj}
\underset{\vect{c} \in \mathcal{C}}{\arg\min}\; f(\vect{c})
\end{equation}
where $\vect{c} = (o_{1}, o_{2}, \dots, o_{n})$ is a configuration such that $o_{n}$ is a configuration option, which might be binary or enumerate. $\mathcal{C}$ is the configuration space. $f$ denotes measuring the actual system for the performance value of $\vect{c}$, which is time-consuming.




\subsection{Tuning with Performance Requirements}

It is possible that one would prefer the best possible performance, e.g., the smaller the runtime, the better, hence we only need to optimize Equation~\ref{eq:singleobj}. However, there are several reasons in practice that it is not always a desired situation. For example, since configuration tuning is expensive, seeking the best possible performance might consume a significant amount of resources and time for little gain~\cite{DBLP:conf/icdcs/FekryCPRH19}. \revision{Indeed, we have seen tuning that achieves a 2 seconds runtime under around one hour budget, but pushing it towards 1.8 seconds requires another 12 hours to do so: the benefit does not seem worth the cost~\cite{DBLP:conf/icse/ChenChen26}. Some recent works have also revealed this~\cite{DBLP:journals/corr/abs-2506-04509}.}
\revision{Yet, in another case, certain ranges of high latency would simply cause too severe penalties, and hence should certainly be avoided; but since this is not reflected in Equation~\ref{eq:singleobj}, the tuning might consider any latency improvement as beneficial.} As a result, practically one often express a performance requirement with some expectations~\cite{DBLP:journals/tosem/ChenL23a,DBLP:conf/icdcs/FekryCPRH19,DBLP:conf/re/EckhardtVFM16}, e.g., ``\texttt{the runtime shall ideally be 2 seconds}'', which could imply that there is certain tolerance for runtime higher than 2 seconds while anything lower than 2 seconds is fully satisfied, meaning a 2 second runtime is already the preferred optimality, even if a lower runtime is achievable with more tuning budget.


To incorporate the performance requirement, a natural way is to combine it with the tuning objective formulated as:
\begin{equation}
\label{eq:perfobj}
\revision{\underset{\vect{c} \in \mathcal{C}}{\arg\max}\; p(v) \mapsto 0 \leq p(v) \leq 1 \text{ where } v=f(\vect{c})} 
\end{equation}
whereby \revision{$\vect{c}$ denotes a configuration}, $v$ is the measured performance; \revision{we seek to maximize $p(v)$ in which $p$ is a preference mapping function, rather than a constraint, that quantifies and converts the measured performance into a satisfaction score of range $[0,1]$ (e.g., $p(v)=0.7$ means $70\%$ satisfied; $p(v)=1$ and $p(v)=0$ denote fully satisfied and fully unsatisfied, respectively). The mapping is specified and defined by the stakeholders according to their preferences (see Section~\ref{sec:q-req}). Here, we are mainly interested in to what extent the configuration's performance satisfies the requirement. 

Indeed, Chen and Li~\cite{DBLP:journals/tosem/ChenL23a} have proved that, compared with optimizing Equation~\ref{eq:singleobj}, optimizing Equation~\ref{eq:perfobj} is more beneficial if the expectation (e.g., runtime of $20$ seconds) is reasonable. For example, a real world requirement is ``\texttt{must be able to retrieve and display within 20 seconds}'' from the \textit{Criminal Tracking Network and Systems} in the \textsc{PURE} dataset~\cite{pure}, for which one interpretation could be that any higher than $20$ seconds runtime are fully unsatisfied ($p(v)=0$) while anything lower than $20$ seconds receives a satisfaction score in $(0,1]$ via a linearly increasing slope. In this case, tuning for the satisfaction score could enable a good number of configurations to be discriminative while a fair amount of the others are not (i.e., $p(v)=0$), which better balances the exploitation and exploration in the tuning to enforce higher satisfaction.}

\subsection{Motivating Issues}


Unfortunately, optimizing Equation~\ref{eq:perfobj} is not always effective. According to the findings of Chen and Li~\cite{DBLP:journals/tosem/ChenL23a} and our own experiments, we discovered two major issues if the given performance requirement is directly used to guide the tuning:
\begin{itemize}
    \item \textbf{Issue 1:} No pressure to push for convergence. For example, the performance requirement might be too strict, i.e., all configurations explored are fully unsatisfied, hence we do not know what to preserve, leading to weak pressure that causes slow convergence. This makes sense as the developers do not always understand whether a performance expectation is feasible.
    \item \textbf{Issue 2:} Stagnation with premature convergence. The performance requirement might too strongly discriminate the configurations, exacerbating the tuning to be trapped at local optima---the sub-optimal satisfaction never change. Indeed, how exactly the performance requirement impacts the tuning is unknown at requirement elicitation.
\end{itemize}

\begin{figure}[!t]
 \centering
  \begin{subfigure}[t]{0.35\columnwidth}
    \centering
\begin{tikzpicture}
    \begin{axis}[
      width=4cm,
    height=4cm,
     axis x line  = bottom,
    axis y line  = left,
    ymax=1.1,
    ymin=-0.1,
    xmax=1.2,
    y label style={at={(0.2,0.5)}},
     xlabel=$v$ runtime (s),
        ylabel=$p(v)$,
        xtick={0,0.2,0.4},
          ytick={0,1},
         xticklabels={0,2,4},
          yticklabels={0,1},
       legend style={at={(0.5,0.95),font=\fontsize{7}{8}\selectfont},
      anchor=north west,legend columns=1}]

 \addplot[no marks,blue,thick,name path=A] plot coordinates {
(0,1)
(0.2,1)
(0.4,0)
(10,0)
};

 \addplot[only marks,mark=*,draw=black,mark options={fill=green!30}] plot coordinates {
(0.34,0.3)
(0.39,0.05)
(0.45,0)
(0.48,0)
(0.55,0)
(0.6,0)
(0.61,0)
(0.62,0)
(0.63,0)
(0.8,0)
(0.95,0)
(1.1,0)
(1.15,0)
};

       \end{axis}
    
    \end{tikzpicture}
  \subcaption{$p_1$ (no pressure)}
  \end{subfigure}
 ~\hspace{-0.6cm}
  \begin{subfigure}[t]{0.35\columnwidth}
    \centering
\begin{tikzpicture}
    \begin{axis}[
      width=4cm,
    height=4cm,
     axis x line  = bottom,
    axis y line  = left,
    ymax=1.1,
    ymin=-0.1,
    xmax=1.2,
     xlabel=$v$ runtime (s),
      y label style={at={(0.2,0.5)}},
        ylabel=$p(v)$,
        xtick={0,0.2,1},
          ytick={0,1},
         xticklabels={0,2,10},
          yticklabels={0,1},
       legend style={at={(0.5,0.95),font=\fontsize{7}{8}\selectfont},
      anchor=north west,legend columns=1}]

 \addplot[no marks,blue,thick,name path=A] plot coordinates {
(0,1)
(0.2,1)
(1,0)
(10,0)
};

 \addplot[only marks,mark=*,draw=black,mark options={fill=green!30}] plot coordinates {
(0.34,0.825)
(0.39,0.7625)
(0.45,0.6875)
(0.48,0.65)
(0.55,0.5625)
(0.6,0.5)
(0.61,0.4875)
(0.62,0.475)
(0.63,0.4625)
(0.8,0.25)
(0.95,0.0625)
(1.1,0)
(1.15,0)

};

       \end{axis}
    
    \end{tikzpicture}
    \subcaption{$p_2$ (stagnation)}
  \end{subfigure}
 ~\hspace{-0.6cm}
  \begin{subfigure}[t]{0.36\columnwidth}
    \centering
\begin{tikzpicture}
    \begin{axis}[
      width=4cm,
    height=4cm,
     axis x line  = bottom,
    axis y line  = left,
    ymax=1.1,
     ymin=-0.1,
    xmax=1.1,
     xlabel=Budget,
        ylabel=$p(v)$,
         y label style={at={(0.2,0.5)}},
          ytick={0,1},
            xtick={0},
          yticklabels={0,1},
       legend style={at={(0.05,1)},draw=none,fill=none,anchor=north west,legend columns=2}]

 \addplot[no marks,red,thick,dashed] plot coordinates {
(0,0)
(0.1,0)
(0.11,0)
(0.5,0)
(0.6,0.05)
(10,0.05)
};

 \addplot[no marks,blue,thick] plot coordinates {
(0,0)
(0.05,0.4625)
(0.06,0.475)
(0.07,0.4875)
(0.08,0.5)
(0.09,0.5625)
(0.1,0.5625)
(0.11,0.5625)
(10,0.6)
};

\addplot[no marks,blue,dotted,thick] plot coordinates {
(0,0.825)
(10,0.825)
};

\addplot[no marks,red,dotted,thick] plot coordinates {
(0,0.3)
(10,0.3)
};

\addlegendentry{$p_1$}
\addlegendentry{$p_2$}
       \end{axis}

       \node[draw=none] at (1.3,1.6) {\footnotesize Best on $p_2$};
       \node[draw=none] at (1.3,0.55) {\footnotesize Best on $p_1$};

       \draw[->] (2,0.6)--(2.3,0.7);
        \draw[->] (2,1.7)--(2.3,1.8);
    \end{tikzpicture}
   \subcaption{Tuning trajectory}
  \end{subfigure}
  
   \caption{The issues when tuning is directly guided by two alternative performance requirements and their quantification $p_1$ and $p_2$ (for the same system). \revision{$p_1$ in (a) indicates that most configurations cannot be discriminated by the satisfactions (expectation too strict: we probably can only find $p(v)=0$), resulting in limited pressure to steer the tuning toward optimality in (c) when using $p_1$ to guide the tuning. $p_2$ in (b) shows that nearly all configurations can be discriminated by the satisfactions (expectation too relaxed), but this can easily trap the tuning in local optima, causing stagnation that is far from the optimality when using $p_2$ as the objective in (c).}}
   \label{fig:mot}
   \vspace{-0.3cm}
 \end{figure}
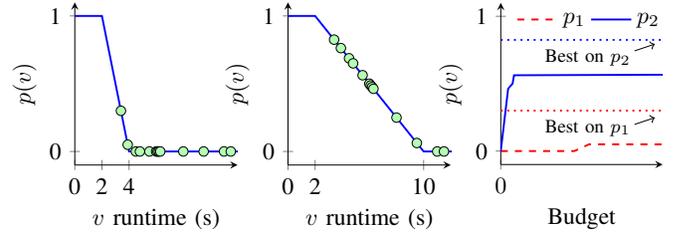

\revision{Figure~\ref{fig:mot}a shows an example for \textbf{Issue 1}, in which the tuning would most likely perform exploration in the landscape with limited guidance, especially under a diversity preservation strategy~\cite{DBLP:journals/jmlr/Bull11}, since it very likely that all configurations that a tuner explored are indistinguishable on satisfying the performance requirement $p_1$ (i.e., $p(v)=0$). In contrast, Figure~\ref{fig:mot}b is a likely case that causes \textbf{Issue 2} where the tuning mostly exploits the performance requirement $p_2$ for guidance since many configurations can be discriminated, but such guidance might be highly strong during the tuning, causing it to quickly converge (prematurely) but easy to stuck at local optima---a typical case of over-emphasizing on exploitation~\cite{DBLP:journals/telo/AthERF21}. It is well known that biasing to either exploration or exploitation is devastating to configuration tuning~\cite{DBLP:journals/corr/abs-2112-07303}, hence the requirement satisfaction of both is struggling to improve, as in Figure~\ref{fig:mot}c.}

\revision{Our motivation is intuitive: if we know what is required by the stakeholders, then using such information to guide the tuning should lead to better results quicker (which is indeed possible~\cite{DBLP:journals/tosem/ChenL23a}). However, due to \textbf{Issues 1-2}, the above is not always true. The key question is \textit{how do we exploit a performance requirement to guide configuration tuning while being able to adapt to its ineffective and/or misleading aspects?} \tuner~is designed precisely to mitigate those.}


\section{Tuning via Co-evolving Performance Requirement and Configurations}
\label{sec:framework}


\tuner~exploits the guidance provided by the given target performance requirement, represented as a proposition\footnote{\revision{Proposition denotes the quantification of performance requirement.}}, while catering for the chance that it can be ineffective or even misleading. Drawing on the co-evolution theory~\cite{DBLP:journals/tec/MaLZTLXZ19}, \tuner~creates a new \textbf{\textit{auxiliary proposition}} ($p_a$) alongside the \textbf{\textit{target proposition}} ($p_t$). $p_t$ would be fixed to serve as the ``golden rule'' but $p_a$ is co-evolved together with the configurations while allowing both the $p_a$ and $p_t$ to guide and evolve the configurations (addressing \textbf{Issues 1-2}). We are not interested in satisfying the auxiliary proposition but merely leverage it to assist the satisfaction of the target one. \tuner~has several phases as in Figure~\ref{fig:overview} and Algorithm~\ref{alg:alg}:

\begin{figure}[!t]
 \centering
\includegraphics[width=\columnwidth]{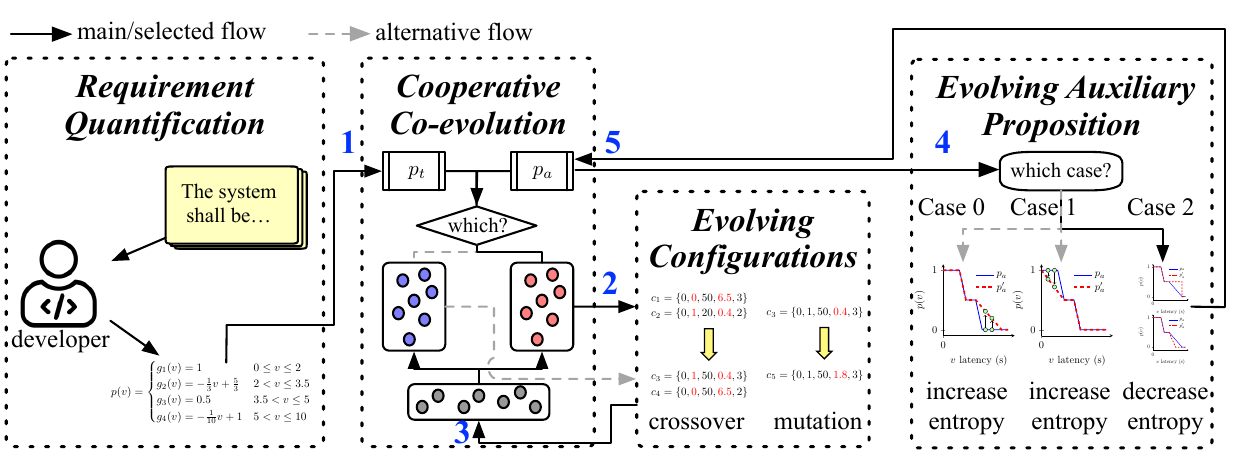} 
   \caption{Architecture overview of \tuner.}
   \label{fig:overview}
   \vspace{-0.3cm}
 \end{figure}


\begin{itemize}
    \item \textbf{Requirements Quantification:} Developers firstly quantify the performance requirement that needs to be satisfied under the framework in \tuner~(Section~\ref{sec:q-req}).
    \item \textbf{Co-evolution with Cooperation:} This determines which proposition to guide the evolution of configurations, and how the newly explored configurations are preserved (Section~\ref{sec:coevo}; lines 8-16 and 27-36).
    \item \textbf{Configuration Evolution:} Configurations are evolved based on the guidance from the selected proposition (Section~\ref{sec:evo-config}; lines 10 and 16).
    \item \textbf{Requirement Evolution:} The auxiliary proposition is evolved depending on whether convergence is too weak or suffering stagnation (Section~\ref{sec:evo-req}; lines 17-26).
\end{itemize}

\tuner~terminates when either a configuration of $p_t(v)=1$ has been found or the budget has been exhausted. 



\begin{algorithm}[t!]
\DontPrintSemicolon
\caption{Pseudo code for \tuner}
\scriptsize
\label{alg:alg}
\textbf{Input:} Budget $B$; population size $n$; stagnation cap $k$; target proposition $p_t$\\
\textbf{Declare:} Auxiliary proposition $p_a$; new auxiliary proposition $p'_a$; temporary auxiliary proposition $p''_a$; set of auxiliary propositions $\mathcal{G}$; $\mathcal{P}_t$ and $\mathcal{P}_a$ are configuration populations for $p_t$ and $p_a$, respectively; the consumed budget $b$; newly evolved configurations set $\mathcal{O}$; configuration $\vect{c}$ and $\vect{c'}$; stagnation counter $i$; the best configuration on $p_t$ from the last iteration $\vect{c}_{pre-best}$\\
\KwOut{The best configuration $\vect{c}_{best}$ on $p_t$}


$p_a = p_t$\\ 
$\mathcal{P}_t, \mathcal{P}_a\leftarrow$ randomly generate $n$ configurations\\
Measure $n$ configurations on the system and fitness via $p_t$ and $p_a$\\
\While{$b + n < B$}{

  $\theta \leftarrow$ compute via Equation~\ref{eq:theta}\\
  \tcc{\revision{Evolve new configurations via standard mating selection, mutation, and crossover at each iteration in Genetic Algorithm via $p_t$ or $p_a$.}}
  \eIf{random $\alpha \in [0,1]< \theta$ or $p_a$ was updated last iteration}{
    $\mathcal{O}\leftarrow$ generate $n$ configurations from $\mathcal{P}_a$ using $p_a$\\
  }{
    $\mathcal{O}\leftarrow$ generate $n$ configurations from $\mathcal{P}_t$ using $p_t$\\
  }

 \tcc{\revision{Preserve promising configurations.}}
  Measure and evaluate configurations in $\mathcal{O}$ under both $p_t$ and $p_a$\\
  $\mathcal{P}_t\leftarrow$ top $n$ configurations from $\mathcal{P}_t \cup \mathcal{O}$ on $p_t$\\
  $\mathcal{P}_a\leftarrow$ top $n$ configurations from $\mathcal{P}_a \cup \mathcal{O}$ on $p_a$\\
  $b = b + n$\\

 \tcc{\revision{Evolve $p_a$ when Case 0 is detected.}}
  \uIf{$\forall p_a(f(\vect{c}))=0 $ and $\forall p_t(f(\vect{c'}))=0, \vect{c} \in \mathcal{P}_a; \vect{c'} \in \mathcal{P}_t$}{
    $p_a \leftarrow$ change $p_a$ by relaxing a boundary point that increases entropy $h$ via Equation~\ref{eq:entropy} and~\ref{eq:zero}
  }
   \tcc{\revision{Evolve $p_a$ when Case 1 is detected.}}
  \uElseIf{$\forall p_a(f(\vect{c}))=1, \vect{c} \in \mathcal{P}_a$}{
    $p_a \leftarrow$ change $p_a$ by tightening a boundary point that increases entropy $h$ via Equation~\ref{eq:entropy} and~\ref{eq:one}
  }
   \tcc{\revision{Evolve $p_a$ when Case 2 is detected.}}
  \ElseIf{$i \geq k$}{
        \While{$\nexists h(f(\vect{c}),p'_a) < h(f(\vect{c}),p_a), p'_a \in \mathcal{G}$ or $|\mathcal{G}|<n$}{
       $\mathcal{G} \leftarrow \mathcal{G} \cup $ generate a new proposition from $p_a$ by changing fragments and/or boundary points\\
       \If{$|\mathcal{G}|>n$}{
         $\mathcal{G} \leftarrow \mathcal{G} - $ $p_a'' \in \mathcal{G}$ of the highest $h$ via Equation~\ref{eq:entropy}
       }
    }
  
    $p_a \leftarrow p_a'' \in \mathcal{G}$ of the lowest $h$ via Equation~\ref{eq:entropy}
  }
  \If{$p_a$ has changed}{
    Reevaluate the fitness of configurations in $\mathcal{P}_a$ on $p_a$
  }
  $\vect{c}_{best}\leftarrow$ the best configuration on $p_t$ from $\mathcal{P}_t \cup \mathcal{P}_a$\\
  
  \If{$\vect{c}_{best}$ is better than $\vect{c}_{pre-best}$ on $p_t$}{
    Set $\vect{c}_{pre-best} = \vect{c}_{best}$ and $i=0$\\
    
     \textbf{if} $p_t(f(\vect{c}_{best}))=1$ \textbf{then} \Return $\vect{c}_{best}$\\
  } \Else {
    $i = i +1$
  }
}

\Return $\vect{c}_{best}$\\
\end{algorithm}


\subsection{Quantifying Performance Requirements}
\label{sec:q-req}

To quantify performance requirements for configuration tuning, we analyze the patterns from the real-world requirements datasets: \textsc{Promise}~\cite{promise}, \textsc{PURE}~\cite{DBLP:conf/re/FerrariSG17,pure}, and \textsc{SRS}~\cite{DBLP:conf/cse/ShaukatNZ18,shaukat}.

\subsubsection{Fragment}


In \tuner, we formally specify performance requirements with fragments as these points essentially represent distinct preferences based on the intervals of metric values. Given an interval $[v_i,v_{i+1}]$ over the value $v$ of a performance metric \revision{(e.g., $[1.5s,3s]$ for latency)}, the fragment and its implied preference of the performance requirement, denoted $\psi$, can be represented via Backus-Naur notations~\cite{DBLP:journals/cacm/Knuth64a}:
\sta{
$\langle\psi\rangle$ ::=  $\mathbf{G}$ $\mathbf{\vert}$ $\mathbf{S}$ $\mathbf{\vert}$ $\mathbf{E}$

$\langle\mathbf{G}\rangle$ ::= $\forall v \in [v_i,v_{i+1}]$, a greater $v$ is preferred at $[s_i,s_{i+1}]$

$\langle\mathbf{S}\rangle$ ::= $\forall v \in [v_i,v_{i+1}]$, a smaller $v$ is preferred at $[s_i,s_{i+1}]$

$\langle\mathbf{E}\rangle$ ::= $\forall v \in [v_i,v_{i+1}]$ is equally preferred at $s_i$
}
\noindent where $s_i$ denotes the satisfaction score for that interval ($s_i \in [0,1]$), which is adapted depending on the preference of the adjacent intervals in a performance requirement. The first two are \textit{\textbf{distinguishable fragments}} while the last is an \textit{\textbf{indistinguishable fragment}}. Clearly, a score of 1 and 0 represent fully satisfied and fully non-satisfied requirement, respectively. Note that while some bounds for a performance metric are clear, e.g., the lower bound for runtime and throughput can only be $0$, in other cases, the bounds may need to be set based on the domain knowledge of developers, e.g., the maximum runtime.



\begin{figure}[!t]
 \centering
  \begin{subfigure}[t]{0.35\columnwidth}
    \centering
\includegraphics[width=\textwidth]{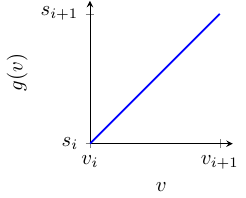}
  \subcaption{$\mathbf{G}$}
  \end{subfigure}
 ~\hspace{-0.6cm}
  \begin{subfigure}[t]{0.35\columnwidth}
    \centering
\includegraphics[width=\textwidth]{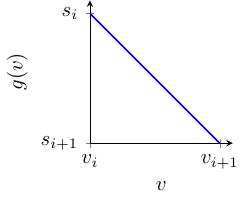}
    \subcaption{$\mathbf{S}$}
  \end{subfigure}
 ~\hspace{-0.6cm}
  \begin{subfigure}[t]{0.35\columnwidth}
    \centering
\includegraphics[width=\textwidth]{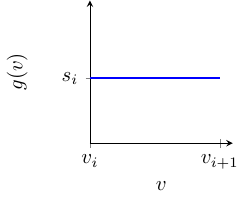}
   \subcaption{$\mathbf{E}$}
  \end{subfigure}
  
    \caption{Fragment types quantified by fuzzy functions $g(v)$ as satisfaction scores. (a): $g(v)=({{s_{i}-s_{i+1}}\over{{v_i-v_{i+1}}}})v + {{s_{i+1}v_i- s_{i}v_{i+1}}\over{{v_i-v_{i+1}}}}$; (b): $g(v)=({{s_{i}-s_{i+1}}\over{{v_i-v_{i+1}}}})v + {{s_{i+1}v_i- s_{i}v_{i+1}}\over{{v_i-v_{i+1}}}}$; (c): $g(v)=s_i$.}
   \label{fig:fragments}
   \vspace{-0.3cm}
 \end{figure}

The satisfaction score of the above fragments can be quantified by a function $g(\cdot)$ using fuzzy logic~\cite{zadeh1988fuzzy} \revision{(e.g., we quantify the extent to which a requirement can be partially satisfied)}. As shown in Figures~\ref{fig:fragments}a and~\ref{fig:fragments}b, since we do not know to what extent a greater (or smaller) $v$ is sufficient, the membership function can be linearly specified as a slop that monotonically increases (or decreases) the satisfaction score gradually from $v_i$ to $v_{i+1}$. Both fragments reach the best and worst satisfaction at the interval bounds\footnote{For maximizing metric, $s_i \leq s_{i+1}$; otherwise, $s_i \geq s_{i+1}$.}. The fragment ``$\forall v \in \theta$ is equally preferred at $s_i$" is a special case of the fuzzy logic, such that all values of the performance metric between $v_i$ and $v_{i+1}$ are equally satisfied at $s_i$ (Figure~\ref{fig:fragments}c).

\subsubsection{Proposition}

In theory, the fragments can be arbitrarily combined, in any number or order, to form a complex yet quantifiable \textit{\textbf{proposition}} for the performance requirement. Here, any two fragments are connected by a \textit{\textbf{boundary point}}. In \tuner, using the Backus-Naur notations, a proposition $p$ of $n$ fragments (with $n-1$ boundary points) is:
\sta{
\centering
$\langle p\rangle$ ::= $\psi$ $\mathbf{\vert}$ $\psi$ \& $p$
}
\noindent in which there are $n$ intervals, i.e., $\{[v_1,v_2], [v_2,v_3]$ $,...,$ $[v_n,v_{n+1}]\}$, and a vector of satisfaction scores, e.g., $\boldsymbol{\overline{s}}=\{[s_1,s_2],s_2,...,$ $[s_{n-1},s_{n}]\}$. Clearly, $p$ is quantified as a piecewise function of several functions $g$ for the fragments. 

In this work, we assume that the above proposition of performance requirements can be elicited and quantified by the developers via formal analysis tools as in prior work~\cite{DBLP:journals/infsof/MethBM13}.


\subsubsection{Example}

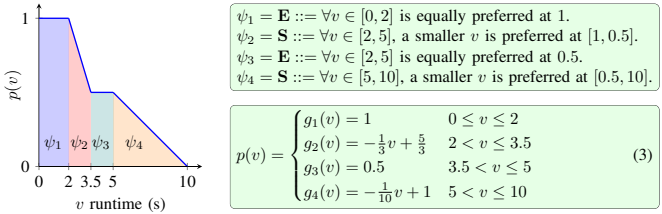
\begin{figure}[!t]
 \centering
\begin{tikzpicture}
    \begin{axis}[
      width=5cm,
    height=5cm,
     axis x line  = bottom,
    axis y line  = left,
    ymax=1.1,
    ymin=0,
    xmax=1.1,
     xlabel=$v$ runtime (s),
        ylabel=$p(v)$,
         y label style={at={(0.2,0.5)}},
        xtick={0,0.2,0.35,0.5,1},
          ytick={0,1},
         xticklabels={0,2,3.5,5,10},
          yticklabels={0,1},
       legend style={at={(0.5,0.95),font=\fontsize{7}{8}\selectfont},
      anchor=north west,legend columns=1}]

 \addplot[no marks,blue,thick,name path=A] plot coordinates {
(0,1)
(0.2,1)
};

 \addplot[no marks,blue,thick,name path=B] plot coordinates {
(0.2,1)
(0.35,0.5)
};

 \addplot[no marks,blue,thick,name path=C] plot coordinates {
(0.35,0.5)
(0.5,0.5)
};

 \addplot[no marks,blue,thick,name path=D] plot coordinates {
(0.5,0.5)
(1,0)
};

 \addplot[no marks,name path=E] plot coordinates {
(0,0)
(0.2,0)
};

 \addplot[no marks,name path=F] plot coordinates {
(0.2,0)
(0.35,0)
};

 \addplot[no marks,name path=G] plot coordinates {
(0.35,0)
(0.5,0)
};

 \addplot[no marks,name path=H] plot coordinates {
(0.5,0)
(1,0)
};

 \addplot[blue!40,opacity=0.5] fill between[of=A and E];
  \addplot[red!40,opacity=0.5] fill between[of=B and F];
  \addplot[teal!40,opacity=0.5] fill between[of=C and G];
  \addplot[orange!40,opacity=0.5] fill between[of=D and H];
%
          
       \end{axis}
         \node at (0.3,0.5) {$\psi_1$};
       \node at (0.85,0.5) {$\psi_2$};
       \node at (1.3,0.5) {$\psi_3$};
          \node at (2,0.5) {$\psi_4$};

  \node at (8.5,2.5) {
 
\begin{minipage}{9cm}
 \begin{innerbox}
$\psi_1 = \mathbf{E} ::=  \forall v \in [0,2]$ is equally preferred at $1$.\\
$\psi_2 = \mathbf{S} ::=  \forall v \in [2,5]$, a smaller $v$ is preferred at $[1,0.5]$.\\
$\psi_3 = \mathbf{E} ::=  \forall v \in [2,5]$ is equally preferred at $0.5$.\\
$\psi_4 = \mathbf{S} ::=  \forall v \in [5,10]$, a smaller $v$ is preferred at $[0.5,10]$.
\end{innerbox}
  \end{minipage}

  }; 

  \node at (8.5,0.2) {
 
\begin{minipage}{9cm}
 \begin{innerbox}
 \vspace{-0.3cm}
  \begin{align}
  \label{eq:exp-rq}
   \hspace{-2.5cm}
p(v) = 
\begin{cases} 
      g_1(v)=1 & 0 \leq v \leq 2 \\
      g_2(v)=-{1 \over 3}v+{5 \over 3} & 2 < v \leq 3.5 \\
      g_3(v)=0.5 & 3.5 < v \leq 5 \\
      g_4(v)=-{1 \over 10}v+1  & 5 < v \leq 10 \\
   \end{cases}
\end{align}
 \end{innerbox}
  \end{minipage}

};  
          
    \end{tikzpicture}
    \caption{Exampled function of a proposition $p(v)$; $v=f(\vect{c})$.}
   \label{fig:exp}
   \vspace{-0.3cm}
 \end{figure}

Consider the performance requirement: ``\texttt{The system should respond in 5 seconds and ideally less than 2 seconds}''. One might interpret that with four fragments and their piecewise function as shown in Figure~\ref{fig:exp} (suppose that we know the runtime $\leq10$ seconds, e.g., due to the timeout setting\footnote{The worst performance metric can often be set based on experience.}). Here, since there is a conflict on the intervals of $v$ for $\psi_2$ and $\psi_3$, we set each of them to share half as $[2,3.5]$ and $[3.5,5]$, respectively.

Our formalization provides a way for developers to reason about and quantify the performance requirements, which can be readily co-evolved with the configurations.

\subsection{Cooperative Co-evolution}
\label{sec:coevo}

\tuner~operates on a cooperative co-evolution~\cite{DBLP:journals/tec/MaLZTLXZ19} between performance requirements and configurations under heterogeneous representations. To that end, we use the auxiliary proposition $p_a$ to assist the target proposition $p_t$, for which we have $p_t=p_a$ initially. During the tuning, $p_t$ is fixed and it is what the tuner needs to satisfy; $p_a$, which will be evolved from $p_t$, guides the tuning when $p_t$ is ineffective or even misleading. These two propositions dynamically and alternatively steer the evolution of the configurations, which, in turn, also impacts how the auxiliary proposition would evolve.

\subsubsection{When to Evolve?}

\revision{Configurations would be evolved at each iteration, which is a standard procedure. For the auxiliary proposition, its evolution might be triggered at an iteration under three cases (with decreasing commonality):}

\begin{itemize}
    \item \revision{The convergence is too weak (too much exploration) due to all explored configurations on the target and auxiliary propositions are fully unsatisfied---over strict proposition.}
    \item \revision{The convergence is too weak (too much exploration) due to all explored configurations on the auxiliary proposition being fully satisfied---the proposition is too relaxed.}
    \item \revision{Suffering stagnation (too much exploitation) where the best configuration on the target proposition has not changed over some iterations.}
\end{itemize}

\subsubsection{How to Evolve?}

\revision{\tuner~evolves configurations following standard Genetic Algorithm~\cite{DBLP:conf/sigsoft/0001L24,k2vtune,DBLP:journals/corr/abs-2112-07303}, i.e., by mating selection, mutation and crossover, to update the configurations according to the target proposition or the auxiliary one if the target becomes misleading. The promising configurations according to the target and auxiliary propositions are preserved separately. The auxiliary proposition is evolved by changing the boundary point and/or switching the fragments ($\mathbf{G}$, $\mathbf{S}$, and $\mathbf{E}$), i.e., replacing the proposition with another similar yet different quantification. The direction and extent of change depend on the aforementioned cases, which are determined by the configurations found and their overall discriminative power, as we will elaborate in Section~\ref{sec:evo-req}. As such, the evolution of the configurations and auxiliary propositions cooperatively influences each other, forming a co-evolution.}

\subsubsection{What Benefits do Co-evolution Bring?}


\revision{This cooperative co-evolution ensures that the evolution of configurations and propositions can positively influence each other. The target proposition still serves a primary role in the tuning when it is effective, and we can leverage the evolving auxiliary proposition to eliminate its negative impact. As such, the benefits of such a co-evolution in \tuner~are two-fold:}

\begin{itemize}
    \item \revision{The useful information embedded in the target proposition can still be leveraged to boost the tuning;}
      \item \revision{while misleading cases (\textbf{Issues 1-2}) are mitigated by the auxiliary proposition co-evolved with the configurations.}
\end{itemize}

\subsubsection{Co-evolution Procedure}

As can be seen from Figure~\ref{fig:overview}, we design two populations of configurations, $\mathcal{P}_t$ and $\mathcal{P}_a$, for evolving under $p_t$ and $p_a$, respectively. These two populations are important to keep track of the evolution via each individual proposition, detecting when the target proposition is misleading while co-evolving the auxiliary one with new configurations generated by the more important proposition. At each iteration, the co-evolution cooperates as follows:

\begin{enumerate}[(a)]
    \item Compute the $\theta$, which reflects the probability for selecting $p_a$ to guide the tuning, as below ($\theta=1$ if $w_a=w_t=0$):
\begin{equation}
\label{eq:theta}
\theta = {{w_a} \over {w_a+w_t}} \text{\phantom{aa} s.t. \phantom{a}}
   \begin{aligned}
       &w_a=f_{\mathcal{P}_a\rightarrow p_t} + l_{\mathcal{P}_a\rightarrow p_a}\\
       &w_t=f_{\mathcal{P}_t\rightarrow p_t} + l_{\mathcal{P}_t\rightarrow p_t}
   \end{aligned}  
\end{equation}
    whereby $f_{\mathcal{P}_a\rightarrow p_t}$ and $f_{\mathcal{P}_t\rightarrow p_t}$ respectively denote the average fitness\footnote{From here, we use fitness and satisfaction score interchangeably.} of $\mathcal{P}_a$ and $\mathcal{P}_t$ evaluated by $p_t$. This assesses how well the two propositions help to guide the tuning towards satisfying the target proposition. $l_{\mathcal{P}_a\rightarrow p_a}$ and $l_{\mathcal{P}_t\rightarrow p_t}$ respectively denote the change of the best configuration in $\mathcal{P}_a$ and $\mathcal{P}_t$ with respect to the fitness on $p_a$ and $p_t$. This tracks the tuning progress achieved by the independent guidance from the two propositions. Thus, the more effective proposition that can encourage good progress would be used to guide the tuning. 
    \item To guide the mating selection, new configurations will be evolved based on the fitness evaluated by $p_a$ if a random number $<\theta$ or if the last iteration has evolved $p_a$; $p_t$ would be selected otherwise (Section~\ref{sec:evo-config}). 
    \item After adding the new configurations to both populations, we preserve the top $n$ configurations ($n$ is the population size) from each population using their corresponding proposition. This helps to track the evolution affected by each proposition via their corresponding population.
    \item $p_a$ will be evolved according to different situations of the configurations evolved, see Section~\ref{sec:evo-req}.
\end{enumerate}

This cooperative co-evolution ensures that the evolution of configurations and propositions can positively influence each other. The $p_t$ still serves as a primary role for the tuning while the evolving $p_a$ eliminates its misleading impacts.

\subsection{Evolving Configurations}
\label{sec:evo-config}

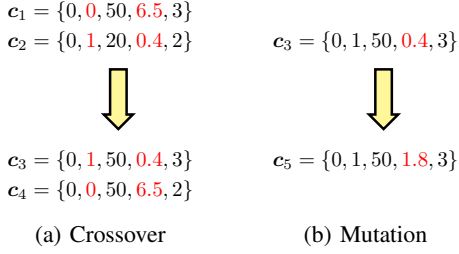
\begin{figure}[!t]
  \centering
 \begin{subfigure}[t]{0.3\columnwidth}
    \centering
\begin{tikzpicture}

\node at (6,2.5) {$\vect{c}_1=\{0,\textcolor{red}{0},50,\textcolor{red}{6.5},3\}$};   

\node at (6,2) {$\vect{c}_2=\{0,\textcolor{red}{1},20,\textcolor{red}{0.4},2\}$};     

\node at (6,0)  {$\vect{c}_3=\{0,\textcolor{red}{1},50,\textcolor{red}{0.4},3\}$}; 

\node at (6,-0.5)  {$\vect{c}_4=\{0,\textcolor{red}{0},50,\textcolor{red}{6.5},2\}$};


  \node[single arrow, draw=black, very thick, fill=yellow!50,
      minimum width = 10pt, single arrow head extend=3pt,
      minimum height=10mm,
      rotate=-90] at (6.3,1.1) {};
    
    \end{tikzpicture}
  \subcaption{Crossover}
  \end{subfigure}
 ~\hspace{0.5cm}
  \begin{subfigure}[t]{0.3\columnwidth}
    \centering
\begin{tikzpicture}

\node at (6,2.5) {\phantom{$\vect{c}_1=\{0,1,50,\textcolor{red}{0.4},3\}$}}; 

\node at (6,2) {$\vect{c}_3=\{0,1,50,\textcolor{red}{0.4},3\}$};     

\node at (6,0)  {$\vect{c}_5=\{0,1,50,\textcolor{red}{1.8},3\}$}; 

\node at (6,-0.5) {\phantom{$\vect{c}_1=\{0,1,50,\textcolor{red}{0.4},3\}$}};


  \node[single arrow, draw=black, very thick, fill=yellow!50,
      minimum width = 10pt, single arrow head extend=3pt,
      minimum height=10mm,
      rotate=-90] at (6.3,1.1) {};
    
    \end{tikzpicture}
    \subcaption{Mutation}
  \end{subfigure}
   \caption{Evolving configurations in \tuner.}
   \label{fig:config}
   \vspace{-0.3cm}
 \end{figure}

The configurations are evolved following the standard procedure in using genetic algorithm for configuration tuning~\cite{k2vtune,DBLP:journals/ase/GerasimouCT18,DBLP:journals/corr/abs-2112-07303,DBLP:conf/wosp/MartensKBR10}. As from Figure~\ref{fig:config}, a configuration is represented as a vector (see Equation~\ref{eq:singleobj}). The reproduction starts by using a binary tournament for mating selection, i.e., randomly choosing two configurations and the one with a better fitness value, which is evaluated by either the target or auxiliary proposition, would be selected (or anyone if they have equal fitness). Two configurations are selected to generate offspring new configurations via the standard random mutation and uniform crossover~\cite{k2vtune,DBLP:journals/corr/abs-2112-07303}. The top $n$ configurations from both the current population and newly generated ones are preserved according to the fitness of the corresponding proposition.

\subsection{Evolving Performance Requirements}
\label{sec:evo-req}

To evolve the auxiliary proposition $p_a$, it is represented as $\{x_1,x_2,...,x_m\}$ where $x_m$ denotes either a fragment or adjustable boundary point (exclude the minimum $v_{min}$ and maximum $v_{max}$ performance value), interleaving each other from left to right. Taking the example in Figure~\ref{fig:exp}, it would become $\{\mathbf{E},2,\mathbf{S},3.5,\mathbf{E},5,\mathbf{S}\}$. The fragment can be switched (e.g., from $\mathbf{S}$ to $\mathbf{E}$) while the boundary value can be changed within the bound of its pieced function; the satisfaction score follows from the left-sided boundary point. $p_a$ would be evolved differently at each iteration depending on three alternative situations to be mitigated, aiming to better assist the target proposition $p_t$ in guiding the tuning, especially when it tends to be useless or misleading: 
\begin{itemize}
    \item \textbf{Case 0:} $\forall p_a(f(\vect{c}))=0 $ and $\forall p_t(f(\vect{c'}))=0, \vect{c} \in \mathcal{P}_a; \vect{c'} \in \mathcal{P}_t$. This means that both the target and auxiliary propositions ($p_t$ and $p_a$) cannot provide sufficient pressure for the tuning to converge, because they are too strict to guide the tuning, making all configurations fully unsatisfied. 
    \item \textbf{Case 1:} \textbf{$\forall p_a(f(\vect{c}))=1, \vect{c} \in \mathcal{P}_a$}. This reflects that the auxiliary proposition $p_a$, which is initially set as $p_a=p_t$ and hence evolved from $p_t$, has become too relaxed and will not provide much convergence pressure if selected, making all configurations fully satisfied. 
    \item \textbf{Case 2:} The best $\vect{c}$ on $p_t$ from both $\mathcal{P}_t$ and $\mathcal{P}_a$ has not changed in $k$ iterations (we use $k$=3 which is the reasonable setting). This implies that although $p_t$ and/or $p_a$ might still offer good search pressure for the tuning, the pressure could be too strong, causing the tuning traps at local optima---a typical stagnation~\cite{DBLP:journals/jmlr/Bull11,DBLP:journals/telo/AthERF21}.
\end{itemize}

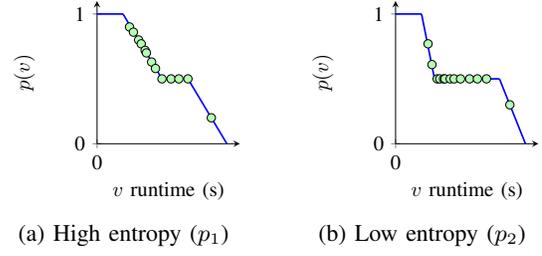
\begin{figure}[!t]
 \centering
  \begin{subfigure}[t]{0.35\columnwidth}
    \centering
\begin{tikzpicture}
    \begin{axis}[
      width=4cm,
    height=4cm,
     axis x line  = bottom,
    axis y line  = left,
    ymax=1.1,
    ymin=0,
    xmax=1.1,
     xlabel=$v$ runtime (s),
        ylabel=$p(v)$,
       xtick={0},
          ytick={0,1},
         xticklabels={0},
          yticklabels={0,1},
       legend style={at={(0.5,0.95),font=\fontsize{7}{8}\selectfont},
      anchor=north west,legend columns=1}]

 \addplot[no marks,blue,thick,name path=A] plot coordinates {
(0,1)
(0.2,1)
};

 \addplot[no marks,blue,thick,name path=B] plot coordinates {
(0.2,1)
(0.5,0.5)
};

 \addplot[no marks,blue,thick,name path=C] plot coordinates {
(0.5,0.5)
(0.65,0.5)
};

 \addplot[no marks,blue,thick,name path=D] plot coordinates {
(0.7,0.5)
(1,0)
};

 \addplot[only marks,mark=*,draw=black,mark options={fill=green!30}] plot coordinates {
(0.25,0.9)
(0.28,0.86)
(0.32,0.8)
(0.34,0.77)
(0.37,0.72)
(0.38,0.7)
(0.42,0.63)
(0.45,0.58)
(0.5,0.5)
(0.57,0.5)
(0.63,0.5)
(0.7,0.5)
(0.88,0.2)

};

       \end{axis}
    
    \end{tikzpicture}
  \subcaption{High entropy ($p_1$)}
  \end{subfigure}
 ~\hspace{0.5cm}
  \begin{subfigure}[t]{0.35\columnwidth}
    \centering
\begin{tikzpicture}
    \begin{axis}[
      width=4cm,
    height=4cm,
     axis x line  = bottom,
    axis y line  = left,
    ymax=1.1,
    ymin=0,
    xmax=1.1,
     xlabel=$v$ runtime (s),
        ylabel=$p(v)$,
       xtick={0},
          ytick={0,1},
         xticklabels={0},
          yticklabels={0,1},
       legend style={at={(0.5,0.95),font=\fontsize{7}{8}\selectfont},
      anchor=north west,legend columns=1}]

 \addplot[no marks,blue,thick,name path=A] plot coordinates {
(0,1)
(0.2,1)
};

 \addplot[no marks,blue,thick,name path=B] plot coordinates {
(0.2,1)
(0.3,0.5)
};

 \addplot[no marks,blue,thick,name path=C] plot coordinates {
(0.3,0.5)
(0.8,0.5)
};

 \addplot[no marks,blue,thick,name path=D] plot coordinates {
(0.8,0.5)
(1,0)
};

 \addplot[only marks,mark=*,draw=black,mark options={fill=green!30}] plot coordinates {
(0.25,0.77)
(0.28,0.61)
(0.32,0.5)
(0.34,0.5)
(0.37,0.5)
(0.38,0.5)
(0.42,0.5)
(0.45,0.5)
(0.5,0.5)
(0.57,0.5)
(0.63,0.5)
(0.7,0.5)
(0.88,0.3)

};

       \end{axis}
    
    \end{tikzpicture}
    \subcaption{Low entropy ($p_2$)}
  \end{subfigure}
  
   \caption{\revision{A proposition $p_1$ with high entropy in (a) and another $p_2$ with low entropy in (b)} for the same set of configurations.}
   \label{fig:en}
   \vspace{-0.3cm}
 \end{figure}

The key cause of the above cases is related to the ability to discriminate the configurations through the (auxiliary) proposition, which is a key to comparing and evolving the better or worse configuration. In \tuner, we leverage differential entropy to quantify such a discriminative power achieved by a proposition $p$ over the current configurations $\vect{c}$ in $\mathcal{P}_a$, \revision{whose (continuous) performance value $v=f(\vect{c})$ is converted into the continuous satisfaction score $p(v)\in[0,1]$:} 
\begin{equation}
\label{eq:entropy}
\revision{h(v,p) = -\int_{\vect{c} \in \mathcal{P}_a} \beta(p(v)) \log \beta(p(v)) dv}
\end{equation}
where $\beta$ is the probability density function of the satisfactions from $p$ for $\vect{c}$ via Kernel Density Estimation (KDE)~\cite{terrell1992variable}.

As we can see from Figure~\ref{fig:en}, in general, differential entropy measures the ``spread'', or uncertainty, of the satisfactions over certain configurations evaluated by a proposition. When $h$ is high (Figure~\ref{fig:en}a), most part of the satisfaction distribution would have more deviated/fluctuated values (hence uncertain), making it easier to distinguish good from poor configurations, which strengthens the ability of discrimination. Such strong guidance might exacerbate the chance of trapping at local optima in \textbf{Issue 2}. Conversely, when $h$ is low (Figure~\ref{fig:en}b), configurations in the distribution tend to have the same, or highly similar, satisfaction scores, hence it is harder to state which configurations are clearly better/worse, weakening the discriminative power. This might cause \textbf{Issue 1} in the tuning. In what follows, we will delineate how we leverage differential entropy to steer the evolution of auxiliary proposition.

\subsubsection{Encouraging Convergence}

The issue behind \textbf{Case 0} and \textbf{Case 1} is that they have no discrimination ability on the configurations, hence losing the pressure for convergence. Thus, the goal of evolving the $p_a$ in \tuner~is to strengthen the discrimination by increasing the differential entropy of the current auxiliary proposition, encouraging exploitation.

\begin{figure}[!t]
 \centering
  \begin{subfigure}[t]{0.35\columnwidth}
    \centering
\begin{tikzpicture}
    \begin{axis}[
      width=4cm,
    height=4cm,
     axis x line  = bottom,
    axis y line  = left,
    ymax=1.1,
     ymin=-0.1,
    xmax=1.1,
    xlabel=$v$ runtime (s),
        ylabel=$p(v)$,
          ytick={0,1},
           xtick={0},
          yticklabels={0,1},
       legend style={at={(0.4,0.95)},draw=none,anchor=north west,legend columns=1}]

 \addplot[no marks,blue,thick] plot coordinates {
(0,1)
(0.25,1)
(0.35,0.5)
(0.5,0.5)
(0.6,0)
(1,0)
};

 \addplot[no marks,red,ultra thick,dashed] plot coordinates {
(0,1)
(0.25,1)
(0.35,0.5)
(0.5,0.5)
(0.9,0)
(1,0)
};

\addplot[only marks,mark=*,draw=black,mark options={fill=green!30}] plot coordinates {
(0.65,0)
(0.75,0)

(0.65,0.31)
(0.75,0.195)

};


\addlegendentry{$p_a$}
\addlegendentry{$p'_a$}

       \end{axis}

\draw[->] (1.645,0.27) -- (1.645,0.5);
\draw[->] (1.425,0.27) -- (1.425,0.7);
    
    \end{tikzpicture}
  \subcaption{Case 0, \revision{$p_a$ to $p'_a$}}
  \end{subfigure}
 ~\hspace{0.5cm}
  \begin{subfigure}[t]{0.35\columnwidth}
    \centering
\begin{tikzpicture}
    \begin{axis}[
      width=4cm,
    height=4cm,
     axis x line  = bottom,
    axis y line  = left,
    ymax=1.1,
     ymin=-0.1,
    xmax=1.1,
    xlabel=$v$ runtime (s),
        ylabel=$p(v)$,
          ytick={0,1},
          xtick={0},
          yticklabels={0,1},
       legend style={at={(0.4,0.95)},draw=none,anchor=north west,legend columns=1}]

 \addplot[no marks,blue,thick] plot coordinates {
(0,1)
(0.25,1)
(0.35,0.5)
(0.5,0.5)
(0.6,0)
(1,0)
};

 \addplot[no marks,red,ultra thick,dashed] plot coordinates {
(0,1)
(0.35,0.5)
(0.5,0.5)
(0.6,0)
(1,0)
};

\addplot[only marks,mark=*,draw=black,mark options={fill=green!30}] plot coordinates {
(0.1,1)
(0.2,1)

(0.1,0.85)
(0.2,0.72)

};


\addlegendentry{$p_a$}
\addlegendentry{$p'_a$}

       \end{axis}

       \draw[->] (0.22,2.14) -- (0.22,2);
 \draw[->] (0.44,2.14) -- (0.44,1.75);
    
    \end{tikzpicture}
    \subcaption{Case 1, \revision{$p_a$ to $p'_a$}}
  \end{subfigure}
  
    \caption{Evolving the auxiliary proposition for Case 0 and 1.}
   \label{fig:case1-2}
   \vspace{-0.3cm}
 \end{figure}
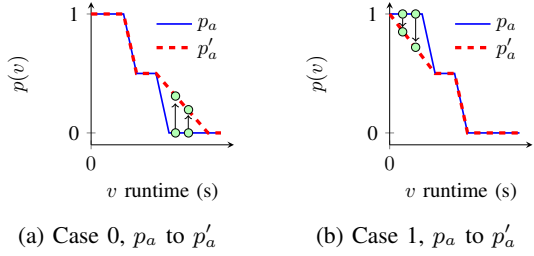

Suppose that the performance objective has been converted to minimization, to evolve a new auxiliary proposition $p'_a$ under \textbf{Case 0}, \tuner~finds the first fragment from the right that is distinguishable and move its right boundary point rightward by $\Delta$ times, where $\Delta$ is randomly chosen from $[0.5,1]$: $b_{right} = \min\{b_{right} + b_{right} \times \Delta, v_{max}\}$ such that $v_{min}$ is the maximum value of the performance metric, if any. The process stops when the current set of configurations $\vect{c}$ has $h(v=f(\vect{c}),p'_a) > h(v=f(\vect{c}),p_a)$: most previously fully unsatisfied configurations might become partially satisfied with varying satisfactions. $p_a$ and $p'_a$ will have:
\begin{equation}
\label{eq:zero}
\int_{v_{min}}^{v_{max}} {p_a}(v)\,dv < \int_{v_{min}}^{v_{max}} {p'_a}(v)\,dv \Rightarrow p_a \sqsubseteq p'_a
\end{equation}
From the example in Figure~\ref{fig:case1-2}a, this means that the area of $p'_a$ will be larger than that of $p_a$, i.e., we have relaxed the auxiliary proposition of the performance requirement. 


Similarly, to evolve a new auxiliary proposition $p'_a$ under \textbf{Case 1}, \tuner~finds the first fragment from the left that is distinguishable and move its left boundary point $b_{left}$ leftward by $\Delta$ times, where $\Delta$ is randomly chosen from $[0.5,1]$: $b_{left} = \max\{b_{left} - b_{left} \times \Delta, v_{min}\}$ such that $v_{min}$ is the minimum value of the performance metric, if any. Again, the process stops when the current set of configurations $\vect{c}$ has $h(v=f(\vect{c}),p'_a) > h(v=f(\vect{c}),p_a)$: most previously fully satisfied configurations might become partially satisfied with varying satisfactions. In this way, $p_a$ and $p'_a$ will meet:
\begin{equation}
\label{eq:one}
\int_{v_{min}}^{v_{max}} {p_a}(v)\,dv > \int_{v_{min}}^{v_{max}} {p'_a}(v)\,dv \Rightarrow p_a \sqsupseteq p'_a
\end{equation}
From Figure~\ref{fig:case1-2}b, in this case, it means that the area of $p'_a$ will be smaller than that of $p_a$, i.e., we have tightened the auxiliary proposition of the performance requirement. 


\subsubsection{Overcoming Stagnation and Local Optima}

\textbf{Case 2} differs from the other two in the sense that the propositions can have strong pressure for convergence by discriminating the configurations, but the tuning might be struggling to escape from some local optima~\cite{DBLP:journals/jmlr/Bull11,DBLP:journals/telo/AthERF21}. To mitigate that, \tuner~seeks to evolve a new auxiliary proposition $p_a$ that exhibits much lower differential entropy, as it is known that weakening the ability to discriminate the configurations is the key to overcome local optima~\cite{DBLP:journals/corr/abs-2112-07303} by encouraging exploration.

Since there is no clear direction to evolve $p_a$, we follow a random search strategy to mutate $p_a$ in the space of possible propositions, guided by differential entropy using all configurations in population $\mathcal{P}_a$ (lines 21-26 in Algorithm~\ref{alg:alg}):

\begin{enumerate}[(a)]
    \item Mutate a new auxiliary proposition $p'_a$ from $p_a$ by randomly changing the fragments and/or boundary points.
     \item Get the entropy $h$ for the generated propositions against all configurations from $\mathcal{P}_a$ by KDE and Equation~\ref{eq:entropy}.
    \item Repeat from (a) until the required number of propositions has been met and at least one new proposition has lower entropy $h$ than the current auxiliary proposition $p_a$.
    \item The proposition with the lowest $h$ is the new $p_a$.
\end{enumerate}

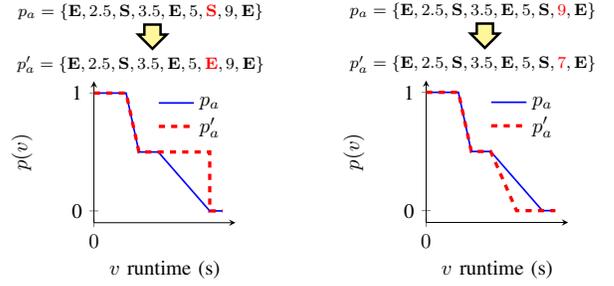
\begin{figure}[!t]
 \centering
  \begin{subfigure}[t]{0.4\columnwidth}
    \centering
\begin{tikzpicture}
    \begin{axis}[
      width=4cm,
    height=4cm,
     axis x line  = bottom,
    axis y line  = left,
    ymax=1.1,
     ymin=-0.1,
    xmax=1.1,
    xlabel=$v$ runtime (s),
        ylabel=$p(v)$,
          ytick={0,1},
           xtick={0},
          yticklabels={0,1},
       legend style={at={(0.4,0.95)},draw=none,anchor=north west,legend columns=1}]

 \addplot[no marks,blue,thick] plot coordinates {
(0,1)
(0.25,1)
(0.35,0.5)
(0.5,0.5)
(0.9,0)
(1,0)
};

 \addplot[no marks,red,ultra thick,dashed] plot coordinates {
(0,1)
(0.25,1)
(0.35,0.5)
(0.5,0.5)
(0.9,0.5)
(0.9,0)
(1,0)
};


\addlegendentry{$p_a$}
\addlegendentry{$p'_a$}

       \end{axis}

  \node at (0.8,3.6) {\footnotesize$p_a=\{\mathbf{E},2.5,\mathbf{S},3.5,\mathbf{E},5,\textcolor{red}{\mathbf{S}},9,\mathbf{E}\}$};     

\node at (0.8,2.7) {\footnotesize$p'_a=\{\mathbf{E},2.5,\mathbf{S},3.5,\mathbf{E},5,\textcolor{red}{\mathbf{E}},9,\mathbf{E}\}$};


  \node[single arrow, draw=black, very thick, fill=yellow!50,
      minimum width = 10pt, single arrow head extend=3pt,
      minimum height=3mm,
      rotate=-90] at (1,3.25) {};
    
    \end{tikzpicture}
  \subcaption{Switching fragment}
  \end{subfigure}
 ~\hspace{0.5cm}
  \begin{subfigure}[t]{0.4\columnwidth}
    \centering
\begin{tikzpicture}
    \begin{axis}[
      width=4cm,
    height=4cm,
     axis x line  = bottom,
    axis y line  = left,
    ymax=1.1,
     ymin=-0.1,
    xmax=1.1,
    xlabel=$v$ runtime (s),
        ylabel=$p(v)$,
          ytick={0,1},
           xtick={0},
          yticklabels={0,1},
       legend style={at={(0.4,0.95)},draw=none,anchor=north west,legend columns=1}]

 \addplot[no marks,blue,thick] plot coordinates {
(0,1)
(0.25,1)
(0.35,0.5)
(0.5,0.5)
(0.9,0)
(1,0)
};

 \addplot[no marks,red,ultra thick,dashed] plot coordinates {
(0,1)
(0.25,1)
(0.35,0.5)
(0.5,0.5)
(0.7,0)
(1,0)
};


\addlegendentry{$p_a$}
\addlegendentry{$p'_a$}

       \end{axis}

       \node at (0.8,3.6) {\footnotesize$p_a=\{\mathbf{E},2.5,\mathbf{S},3.5,\mathbf{E},5,{\mathbf{S}},\textcolor{red}{9},\mathbf{E}\}$};     

\node at (0.8,2.7) {\footnotesize$p'_a=\{\mathbf{E},2.5,\mathbf{S},3.5,\mathbf{E},5,{\mathbf{S}},\textcolor{red}{7},\mathbf{E}\}$};


  \node[single arrow, draw=black, very thick, fill=yellow!50,
      minimum width = 10pt, single arrow head extend=3pt,
      minimum height=3mm,
      rotate=-90] at (1,3.25) {};
    \end{tikzpicture}
    \subcaption{Moving boundary point}
  \end{subfigure}
   \caption{Evolving auxiliary proposition for Case 2.}
   \label{fig:case3}
   \vspace{-0.1cm}
 \end{figure}

As in Figure~\ref{fig:case3}, the mutations ensure that we do not only change the boundary points (Figure~\ref{fig:case3}b), but can also switch the fragments (Figure~\ref{fig:case3}a)---the proposition can either be overall relaxed or tightened, as long as the differential entropy over the current configurations is minimized, weakening the discriminative power of tuning to jump out of local optima.

\section{Experiment Setup}
\label{sec:experiment_setup}

\begin{table}[t!]
\centering

\setlength{\tabcolsep}{0.8mm}
\footnotesize
\caption{Real-world configurable systems studied. $|\mathcal{B}|$ and $|\mathcal{N}|$ denote the number of binary and numeric/enumerate options, respectively; $\mathcal{S}_{space}$ denotes the configuration space.}
\label{tb: systems}

\begin{adjustbox}{width=\columnwidth,center}
\begin{tabular}{llllllrrl}
\toprule
\textbf{System}  & \textbf{Benchmark} & \textbf{Domain} & \textbf{Lang.} & \textbf{Perf. Metric} & \textbf{$|\mathcal{B}|$/$|\mathcal{N}|$} & \textbf{$\mathcal{S}_{space}$} & \textbf{Ref.} \\
\midrule
\textsc{7z}~\cite{7z}  & A 3 GB directory & File Compressor & C$++$ & Runtime (ms) & 11/3 & 4.39$\times10^5$ & \cite{DBLP:conf/icse/WeberKSAS23} \\
\textsc{Kanzi}~\cite{kanzi}  & Default benchmark & File Compressor & Java & Runtime (ms) & 31/0 & 5.36$\times10^8$ & \cite{chen2025accuracy} \\
\textsc{ExaStencils}~\cite{exastencils}  & Default benchmark & Code Generator & Scala & Runtime (ms) & 7/5  &  6.55$\times10^5$ & \cite{DBLP:conf/icse/WeberKSAS23} \\ 
\textsc{Apache}~\cite{apache}  & Web server benchmark & Web Server & C & Throughput (req/s) & 14/2 & 3.28$\times10^4$ & \cite{chen2025accuracy} \\
\textsc{SQLite}~\cite{sqlite}  & Default benchmark & Database Engine & Various & Latency (ms) & 39/0 & 5.50$\times10^{11}$ & \cite{chen2025accuracy} \\
\textsc{DConvert}~\cite{muhlbauer2023analyzing} & Default images & Image Scaling & Java & Runtime (s) & 17/1 &  2.62$\times10^5$ & \cite{DBLP:conf/icse/MuhlbauerSKDAS23} \\
\textsc{DeepArch}~\cite{deeparch}  & UCR Archive dataset & AI Tool & Python & Runtime (min) & 12/0  & 4.10$\times10^3$ & \cite{DBLP:conf/sigsoft/JamshidiVKS18} \\ 
\textsc{Jump3r}~\cite{jump3r}  & Jump3r codebase & Static Analysis & Java & Runtime (ms) & 37/0 & 6.87$\times10^{10}$ & \cite{chen2025accuracy} \\
\textsc{HSMGP}~\cite{hsmgp}  & V-cycle solver bench & Multigrid Solver & C$++$ & Runtime (ms) & 11/3 & 1.00$\times10^5$ & \cite{chen2025accuracy} \\


\bottomrule
\end{tabular}
\end{adjustbox}
\vspace{-0.3cm}
\end{table}


To assess \tuner, we study four research questions (RQs):

\begin{itemize}
  \item \textbf{RQ1:} \revision{How can the co-evolution help in \tuner}?
  \item \textbf{RQ2:} How does \Model~perform against the others?
  \item \textbf{RQ3:} What is the contribution of the key component to \Model's performance?
  \item \textbf{RQ4:} What is the sensitivity of \Model~to $k$?
\end{itemize}


All experiments were implemented in Python and performed on a server with Intel(R) CPU (224 cores) and 500GB RAM.


\subsection{Subject Configurable Systems}

To ensure a fair evaluation, we use systems and their performance data measured under real-world benchmarks. We prioritize systems that have been widely used in the literature:
\begin{itemize}
  \item For multiple versions of a system, we use the one with more options to ensure relevance and complexity.

  \item We exclude systems that lack consistent benchmarking protocols or have been used by $<2$ study for reliability.

\end{itemize}

Table~\ref{tb: systems} shows all the systems studied, covering a wide range of domains/types, scales, metrics, and programming languages. For simplicity, we convert all maximizing performance metrics into minimization via additive inversion.


\subsection{Compared Tuners}

We consider several state-of-the-art tuners to fulfil the RQs, including the widely used Genetic Algorithm (\texttt{GA}) based tuner~\cite{k2vtune,DBLP:journals/ase/GerasimouCT18,DBLP:journals/corr/abs-2112-07303,DBLP:conf/wosp/MartensKBR10} (i.e., \tuner~without co-evolving the performance requirement) and those rely on the model-based Bayesian Optimization, i.e., \texttt{HEBO}~\cite{cowen2022hebo}, \texttt{Flash}~\cite{DBLP:journals/tse/Nair0MSA20}, and \texttt{SMAC}~\cite{SMAC}, \revision{together with the most recent ones: \texttt{TurBO}~\cite{DBLP:conf/nips/ErikssonPGTP19} and \texttt{Bounce}~\cite{DBLP:conf/nips/PapenmeierNP23}. We omitted those that have been discarded, e.g., \texttt{BaxUS}~\cite{DBLP:conf/nips/PapenmeierNP22} has been upgraded to \texttt{Bounce}.} 

\revision{We compare two variants of each compared tuner $X$ (including all variants of Bayesian Optimization):} 

\begin{itemize}
    \item \revision{$X_r$: The origin that tunes without knowing the performance requirements, i.e., optimizing Equation~\ref{eq:singleobj}.} 
    \item \revision{$X_p$: The tuning is guided by the requirement---the satisfaction scores of configurations' actual/predicted performance (via $p_t$)---instead of the raw performance in acquisition/selection, i.e., optimizing Equation~\ref{eq:perfobj}.} 
\end{itemize}

\revision{Their evaluations are the same: through the given target proposition $p_t$, we compare the satisfaction score of the performance of the returned configuration from the tuner.} 

\revision{Note that we do not consider multi-fidelity tuners~\cite{DBLP:conf/ijcai/AwadMH21,DBLP:conf/nips/MallikBHSJLNH23,DBLP:conf/icml/FalknerKH18}, because they leverage a well-defined fidelity for AutoML, i.e., using more training data will have higher-fidelity accuracy but is more costly, which is unclear for configurable systems, e.g., for an image rescaling system, it is unclear how the images can be changed to influence the system performance/cost.}




\subsection{Requirement, Budget, and Parameter Settings}

\revision{Since we found no requirement datasets that are specifically related to the studied systems under tuning, in this work, we synthetically generate requirements in the evaluation according to the systems. However, the generated ones are not arbitrary, but rather they are instantiated based on patterns extracted from real-world requirements datasets with $100+$ performance requirements~\cite{promise,DBLP:conf/cse/ShaukatNZ18,DBLP:conf/re/FerrariSG17}, including \textsc{Promise}~\cite{promise}, \textsc{PURE}~\cite{DBLP:conf/re/FerrariSG17,pure}, and \textsc{SRS}~\cite{DBLP:conf/cse/ShaukatNZ18,shaukat}, and tailor them to be system specific:}

\begin{itemize}
    \item \revision{We generate each requirement with five fragments, which is the most complex case identified from those datasets.
    \item For each requirement, we randomly and proportionally assign the fragment ($\mathbf{G}$, $\mathbf{S}$, and $\mathbf{E}$) according to the datasets, while ensuring monotonic satisfaction, e.g., a lower latency would never be less preferred than a higher value; they are at most equally preferred/non-preferred.}
    \item \revision{Set the expectation/boundary point according to the empirical ranges for the corresponding systems/metrics.}
\end{itemize}

\revision{The above is also a similar strategy adopted by Chen and Li~\cite{DBLP:journals/tosem/ChenL23a}.} For each system, we generate the target proposition $p_t$ with the procedure below:



\begin{enumerate}[(a)]
    \item Randomly generate a $p_t$ with five fragments\footnote{In fact, with random generation we can have two fragments joined by a boundary point, where both of them have monotonically decreasing, increasing or consistent satisfactions. This reduces the number of fragments as the two adjacent fragments are quantified in the same way as if there are no boundary points in between, hence also emulating the cases of simpler propositions.}.
    \item Randomly and finely replace the fragments and/or change the boundary points such that $d\%$ of the configurations fully or partially satisfy $p_t$, where $d \in \{0.1\%,1\%,5\%,20\%,50\%,90\%\}$. A lower $d\%$ implies a more strict requirement, which is harder to satisfy.
    \item Repeat (b) until all \% caps are covered.
    \item Repeat from (a) for three times.
\end{enumerate}


In total, there are three types of requirements ($p_{t,1}$, $p_{t,2}$, and $p_{t,3}$) $\times$ six levels of satisfiability $=18$ performance requirements per system. As such, different systems would be given target performance requirements of diverse difficulties, shapes, and implications to the tuning if used therein. 


We use the number of system measurements as tuning budget which is robust to the noises of the implementation and hardware~\cite{DBLP:journals/tse/Nair0MSA20,chen2025accuracy,DBLP:journals/corr/abs-2112-07303}. We set a maximum budget of 300 since it is a most common setting from the literature\footnote{For fair comparisons, we do not stop any tuner even if $p_t=1$ has been achieved before all budget has been used.}~\cite{chen2025accuracy,DBLP:journals/corr/abs-2112-07303}. For \tuner~and \texttt{GA}, we use 30 generations and a population size of 10, which is again the default setting for configuration tuning~\cite{DBLP:journals/corr/abs-2112-07303}. For mitigating stagnation in \tuner, we set $k=3$ and this will be evaluated in \textbf{RQ4}. The mutation and crossover rate for configuration is 0.1 and 0.9, respectively, as with existing work~\cite{DBLP:journals/corr/abs-2112-07303}. For model-based tuners, the initial sample size is 10 and all other settings are left as default~\cite{DBLP:journals/tse/Nair0MSA20}.



All experiments run 30 repeats with different random seeds. 



\subsection{Metrics and Statistical Test}


The metric to be evaluated is simply the satisfaction score of the given target performance requirement $p_t$.

To compare multiple tuners over 30 runs, we adopt the \revision{non-parametric} Scott-Knott ESD~\cite{DBLP:journals/tse/Tantithamthavorn19} as the statistical test, which ranks tuners based on their average satisfaction and partitions them into statistically distinct groups by maximizing inter-group variance, e.g., if three tuners $A$, $B$, and $C$ are evaluated, the ranking result may yield $\{A, B\}$ as 1 and $\{C\}$ as 2, thus $A$ and $B$ perform similarly, but both outperform $C$.

Compared with other multi-groups tests with post-hoc corrections~\cite{mchugh2011multiple,mckight2010kruskal}, Scott-Knott ESD overcomes the confounding factor of overlapping groups~\cite{DBLP:journals/tse/Tantithamthavorn19} with interpretable results. 





\section{Evaluation}

\label{sec:results}

\subsection{Importance of Co-evolution for Tuning}



\begin{table}[t!]
  \setlength{\tabcolsep}{0.8mm}
  \caption{Comparing \tuner~with \texttt{GA} on a budget of 300 over 30 runs. Each cell (except the last row) reports the mean$\pm$standard deviation of the satisfaction (Scott-Knott ESD rank). A \setlength{\fboxsep}{1.5pt}\colorbox{green!20}{green cell} denotes the best ranked tuner for a case.}

  \label{tab:rq1}

  \begin{adjustbox}{max width=\columnwidth,center}
    \sisetup{table-format=1.3}
    \centering
    \begin{tabular}{lllll|lll|lll}
      \toprule
      \multirow{2}{*}{\textbf{\textbf{${d\%}$}}}
        & \multirow{2}{*}{\textbf{System}}
        & \multicolumn{3}{c|}{$p_{t,1}$}
        & \multicolumn{3}{c|}{$p_{t,2}$}
        & \multicolumn{3}{c}{$p_{t,3}$} \\
      \cmidrule(lr){3-11} 
        & 
        & \multicolumn{1}{c}{\tuner} & \multicolumn{1}{c}{\texttt{GA}$_p$} & \multicolumn{1}{c|}{\texttt{GA}$_r$}
        & \multicolumn{1}{c}{\tuner} & \multicolumn{1}{c}{\texttt{GA}$_p$} & \multicolumn{1}{c|}{\texttt{GA}$_r$}
        & \multicolumn{1}{c}{\tuner} & \multicolumn{1}{c}{\texttt{GA}$_p$} & \multicolumn{1}{c}{\texttt{GA}$_r$} \\
      \midrule

      \multirow{9}{*}{$0.1\%$}
& \textsc{7z} & \cellcolor{green!20}.28$\pm$.34 (1) & .10$\pm$.25 (2) & \cellcolor{green!20}.34$\pm$.37 (1) & .35$\pm$.39 (2) & .09$\pm$.27 (3) & \cellcolor{green!20}.44$\pm$.43 (1) & \cellcolor{green!20}.21$\pm$.33 (1) & .08$\pm$.26 (2) & \cellcolor{green!20}.26$\pm$.35 (1) \\
& \textsc{Kanzi} & \cellcolor{green!20}.00$\pm$.00 (1) & \cellcolor{green!20}.00$\pm$.00 (1) & \cellcolor{green!20}.00$\pm$.00 (1) & \cellcolor{green!20}.01$\pm$.05 (1) & .00$\pm$.00 (2) & .00$\pm$.00 (2) & .00$\pm$.00 (2) & \cellcolor{green!20}.02$\pm$.10 (1) & .00$\pm$.00 (2) \\
& \textsc{ExaStencils} & \cellcolor{green!20}.69$\pm$.46 (1) & .00$\pm$.00 (2) & \cellcolor{green!20}.90$\pm$.30 (1) & .69$\pm$.46 (2) & .03$\pm$.18 (3) & \cellcolor{green!20}.90$\pm$.30 (1) & .62$\pm$.49 (2) & .07$\pm$.25 (3) & \cellcolor{green!20}.83$\pm$.38 (1) \\
& \textsc{Apache} & \cellcolor{green!20}.00$\pm$.01 (1) & .00$\pm$.00 (2) & .00$\pm$.00 (2) & \cellcolor{green!20}.00$\pm$.00 (1) & \cellcolor{green!20}.00$\pm$.00 (1) & \cellcolor{green!20}.00$\pm$.00 (1) & \cellcolor{green!20}.00$\pm$.00 (1) & \cellcolor{green!20}.00$\pm$.00 (1) & \cellcolor{green!20}.00$\pm$.00 (1) \\
& \textsc{SQLite} & .00$\pm$.01 (2) & .00$\pm$.00 (2) & \cellcolor{green!20}.02$\pm$.11 (1) & \cellcolor{green!20}.03$\pm$.18 (1) & \cellcolor{green!20}.03$\pm$.14 (1) & \cellcolor{green!20}.04$\pm$.13 (1) & \cellcolor{green!20}.01$\pm$.03 (1) & .00$\pm$.00 (2) & \cellcolor{green!20}.01$\pm$.06 (1) \\
& \textsc{DConvert} & \cellcolor{green!20}.00$\pm$.00 (1) & \cellcolor{green!20}.00$\pm$.00 (1) & \cellcolor{green!20}.00$\pm$.00 (1) & \cellcolor{green!20}.00$\pm$.00 (1) & \cellcolor{green!20}.00$\pm$.00 (1) & \cellcolor{green!20}.00$\pm$.00 (1) & \cellcolor{green!20}.03$\pm$.18 (1) & .00$\pm$.00 (2) & .00$\pm$.00 (2) \\
& \textsc{DeepArch} & \cellcolor{green!20}.66$\pm$.44 (1) & .24$\pm$.39 (2) & \cellcolor{green!20}.59$\pm$.45 (1) & \cellcolor{green!20}.73$\pm$.40 (1) & .15$\pm$.31 (3) & .62$\pm$.43 (2) & \cellcolor{green!20}.72$\pm$.42 (1) & .26$\pm$.41 (3) & .58$\pm$.45 (2) \\
& \textsc{Jump3r} & \cellcolor{green!20}.00$\pm$.00 (1) & \cellcolor{green!20}.00$\pm$.00 (1) & \cellcolor{green!20}.00$\pm$.00 (1) & \cellcolor{green!20}.00$\pm$.00 (1) & .00$\pm$.00 (2) & .00$\pm$.00 (2) & \cellcolor{green!20}.00$\pm$.00 (1) & .00$\pm$.00 (2) & .00$\pm$.00 (2) \\
& \textsc{HSMGP} & \cellcolor{green!20}.92$\pm$.23 (1) & .14$\pm$.34 (2) & \cellcolor{green!20}.89$\pm$.30 (1) & .75$\pm$.40 (2) & .10$\pm$.30 (3) & \cellcolor{green!20}.84$\pm$.32 (1) & .76$\pm$.42 (2) & .03$\pm$.18 (3) & \cellcolor{green!20}.85$\pm$.33 (1) \\


       \hline       
      
      \multirow{9}{*}{$1\%$}
 & \textsc{7z} & .54$\pm$.39 (2) & \cellcolor{green!20}.66$\pm$.34 (1) & .60$\pm$.37 (2) & .16$\pm$.18 (2) & \cellcolor{green!20}.21$\pm$.18 (1) & \cellcolor{green!20}.23$\pm$.18 (1) & \cellcolor{green!20}.34$\pm$.31 (1) & \cellcolor{green!20}.38$\pm$.30 (1) & \cellcolor{green!20}.36$\pm$.28 (1) \\
& \textsc{Kanzi} & \cellcolor{green!20}.10$\pm$.28 (1) & .06$\pm$.19 (2) & .03$\pm$.14 (2) & .05$\pm$.18 (2) & \cellcolor{green!20}.12$\pm$.28 (1) & \cellcolor{green!20}.09$\pm$.27 (1) & \cellcolor{green!20}.06$\pm$.21 (1) & .02$\pm$.12 (2) & \cellcolor{green!20}.05$\pm$.18 (1) \\
& \textsc{ExaStencils} & \cellcolor{green!20}1.0$\pm$.00 (1) & .76$\pm$.41 (3) & .96$\pm$.15 (2) & .80$\pm$.26 (2) & .52$\pm$.44 (3) & \cellcolor{green!20}.92$\pm$.07 (1) & \cellcolor{green!20}.92$\pm$.07 (1) & .61$\pm$.43 (2) & \cellcolor{green!20}.93$\pm$.06 (1) \\
& \textsc{Apache} & \cellcolor{green!20}.03$\pm$.12 (1) & .00$\pm$.00 (2) & .00$\pm$.00 (2) & \cellcolor{green!20}.01$\pm$.06 (1) & .00$\pm$.00 (2) & .00$\pm$.00 (2) & \cellcolor{green!20}.01$\pm$.07 (1) & .00$\pm$.00 (2) & .00$\pm$.00 (2) \\
& \textsc{SQLite} & \cellcolor{green!20}.20$\pm$.28 (1) & .10$\pm$.24 (2) & \cellcolor{green!20}.19$\pm$.33 (1) & \cellcolor{green!20}.19$\pm$.30 (1) & \cellcolor{green!20}.19$\pm$.34 (1) & .12$\pm$.27 (2) & \cellcolor{green!20}.11$\pm$.23 (1) & .04$\pm$.13 (2) & .05$\pm$.16 (2) \\
& \textsc{DConvert} & \cellcolor{green!20}.41$\pm$.14 (1) & .19$\pm$.21 (3) & .31$\pm$.22 (2) & \cellcolor{green!20}.25$\pm$.14 (1) & .14$\pm$.16 (3) & .18$\pm$.15 (2) & \cellcolor{green!20}.24$\pm$.12 (1) & .07$\pm$.12 (2) & \cellcolor{green!20}.24$\pm$.13 (1) \\
& \textsc{DeepArch} & \cellcolor{green!20}.91$\pm$.21 (1) & .75$\pm$.31 (3) & .83$\pm$.20 (2) & \cellcolor{green!20}.93$\pm$.13 (1) & .64$\pm$.39 (2) & \cellcolor{green!20}.94$\pm$.12 (1) & \cellcolor{green!20}.84$\pm$.26 (1) & \cellcolor{green!20}.79$\pm$.36 (1) & .71$\pm$.31 (2) \\
& \textsc{Jump3r} & \cellcolor{green!20}.08$\pm$.20 (1) & \cellcolor{green!20}.09$\pm$.23 (1) & .00$\pm$.00 (2) & \cellcolor{green!20}.08$\pm$.21 (1) & .02$\pm$.10 (2) & .03$\pm$.15 (2) & \cellcolor{green!20}.04$\pm$.13 (1) & \cellcolor{green!20}.02$\pm$.11 (1) & .00$\pm$.00 (2) \\
& \textsc{HSMGP} & \cellcolor{green!20}.96$\pm$.18 (1) & .58$\pm$.49 (3) & .82$\pm$.38 (2) & \cellcolor{green!20}1.0$\pm$.01 (1) & .48$\pm$.50 (3) & .89$\pm$.30 (2) & \cellcolor{green!20}.93$\pm$.25 (1) & .69$\pm$.46 (3) & .82$\pm$.38 (2) \\

       \hline       
      
      \multirow{9}{*}{$5\%$}
 & \textsc{7z} & \cellcolor{green!20}.72$\pm$.20 (1) & .66$\pm$.22 (2) & .55$\pm$.33 (3) & .62$\pm$.41 (2) & \cellcolor{green!20}.78$\pm$.30 (1) & \cellcolor{green!20}.71$\pm$.36 (1) & \cellcolor{green!20}.59$\pm$.36 (1) & \cellcolor{green!20}.66$\pm$.33 (1) & \cellcolor{green!20}.61$\pm$.37 (1) \\
& \textsc{Kanzi} & \cellcolor{green!20}.24$\pm$.32 (1) & .13$\pm$.28 (2) & .07$\pm$.21 (3) & .14$\pm$.25 (2) & \cellcolor{green!20}.29$\pm$.36 (1) & .17$\pm$.31 (2) & \cellcolor{green!20}.29$\pm$.32 (1) & .18$\pm$.27 (2) & .09$\pm$.21 (3) \\
& \textsc{ExaStencils} & \cellcolor{green!20}.95$\pm$.14 (1) & .92$\pm$.23 (2) & \cellcolor{green!20}.95$\pm$.13 (1) & \cellcolor{green!20}.95$\pm$.07 (1) & .93$\pm$.11 (2) & \cellcolor{green!20}.96$\pm$.05 (1) & \cellcolor{green!20}.91$\pm$.12 (1) & .75$\pm$.31 (3) & .85$\pm$.22 (2) \\
& \textsc{Apache} & \cellcolor{green!20}.13$\pm$.06 (1) & \cellcolor{green!20}.12$\pm$.06 (1) & \cellcolor{green!20}.12$\pm$.07 (1) & \cellcolor{green!20}.34$\pm$.16 (1) & \cellcolor{green!20}.34$\pm$.08 (1) & .28$\pm$.12 (2) & \cellcolor{green!20}.22$\pm$.13 (1) & \cellcolor{green!20}.21$\pm$.08 (1) & .18$\pm$.09 (2) \\
& \textsc{SQLite} & \cellcolor{green!20}.21$\pm$.22 (1) & \cellcolor{green!20}.17$\pm$.21 (1) & .11$\pm$.18 (2) & \cellcolor{green!20}.42$\pm$.39 (1) & \cellcolor{green!20}.36$\pm$.37 (1) & .20$\pm$.31 (2) & \cellcolor{green!20}.23$\pm$.27 (1) & \cellcolor{green!20}.23$\pm$.25 (1) & \cellcolor{green!20}.21$\pm$.30 (1) \\
& \textsc{DConvert} & \cellcolor{green!20}.75$\pm$.21 (1) & \cellcolor{green!20}.72$\pm$.22 (1) & \cellcolor{green!20}.71$\pm$.29 (1) & \cellcolor{green!20}.77$\pm$.16 (1) & \cellcolor{green!20}.78$\pm$.16 (1) & \cellcolor{green!20}.79$\pm$.07 (1) & \cellcolor{green!20}.48$\pm$.08 (1) & .38$\pm$.20 (3) & .44$\pm$.14 (2) \\
& \textsc{DeepArch} & \cellcolor{green!20}.99$\pm$.01 (1) & .96$\pm$.18 (3) & .99$\pm$.02 (2) & \cellcolor{green!20}.99$\pm$.01 (1) & \cellcolor{green!20}.99$\pm$.01 (1) & \cellcolor{green!20}.99$\pm$.01 (1) & \cellcolor{green!20}.99$\pm$.01 (1) & \cellcolor{green!20}.99$\pm$.01 (1) & \cellcolor{green!20}.99$\pm$.01 (1) \\
& \textsc{Jump3r} & \cellcolor{green!20}.15$\pm$.31 (1) & \cellcolor{green!20}.12$\pm$.29 (1) & .08$\pm$.24 (2) & \cellcolor{green!20}.18$\pm$.34 (1) & .09$\pm$.26 (2) & .04$\pm$.17 (3) & \cellcolor{green!20}.08$\pm$.20 (1) & .00$\pm$.01 (2) & \cellcolor{green!20}.08$\pm$.24 (1) \\
& \textsc{HSMGP} & \cellcolor{green!20}1.0$\pm$.00 (1) & .93$\pm$.25 (2) & .90$\pm$.30 (2) & \cellcolor{green!20}1.0$\pm$.00 (1) & .93$\pm$.25 (2) & .86$\pm$.34 (3) & \cellcolor{green!20}.96$\pm$.18 (1) & \cellcolor{green!20}.93$\pm$.25 (1) & .86$\pm$.34 (2) \\

       \hline       
      \multirow{9}{*}{$20\%$}
& \textsc{7z} & \cellcolor{green!20}.91$\pm$.16 (1) & .75$\pm$.33 (3) & .85$\pm$.27 (2) & \cellcolor{green!20}.70$\pm$.15 (1) & \cellcolor{green!20}.70$\pm$.13 (1) & .63$\pm$.18 (2) & .82$\pm$.21 (2) & \cellcolor{green!20}.86$\pm$.15 (1) & .83$\pm$.17 (2) \\
& \textsc{Kanzi} & \cellcolor{green!20}.72$\pm$.29 (1) & .59$\pm$.38 (2) & .32$\pm$.36 (3) & \cellcolor{green!20}.64$\pm$.23 (1) & .50$\pm$.29 (2) & .27$\pm$.30 (3) & \cellcolor{green!20}.37$\pm$.19 (1) & \cellcolor{green!20}.37$\pm$.16 (1) & .26$\pm$.20 (2) \\
& \textsc{ExaStencils} & \cellcolor{green!20}.98$\pm$.05 (1) & \cellcolor{green!20}.99$\pm$.04 (1) & \cellcolor{green!20}.99$\pm$.04 (1) & \cellcolor{green!20}.99$\pm$.04 (1) & .98$\pm$.07 (2) & .98$\pm$.07 (2) & .96$\pm$.07 (2) & .95$\pm$.09 (2) & \cellcolor{green!20}.97$\pm$.04 (1) \\
& \textsc{Apache} & \cellcolor{green!20}.67$\pm$.08 (1) & .62$\pm$.09 (2) & .63$\pm$.10 (2) & \cellcolor{green!20}.14$\pm$.02 (1) & \cellcolor{green!20}.14$\pm$.02 (1) & .13$\pm$.02 (2) & \cellcolor{green!20}.14$\pm$.07 (1) & .13$\pm$.05 (2) & .10$\pm$.04 (3) \\
& \textsc{SQLite} & \cellcolor{green!20}.74$\pm$.26 (1) & .64$\pm$.28 (2) & .48$\pm$.34 (3) & \cellcolor{green!20}.64$\pm$.19 (1) & .52$\pm$.24 (2) & .46$\pm$.26 (3) & \cellcolor{green!20}.24$\pm$.16 (1) & \cellcolor{green!20}.25$\pm$.19 (1) & .18$\pm$.22 (2) \\
& \textsc{DConvert} & \cellcolor{green!20}.91$\pm$.03 (1) & .90$\pm$.05 (2) & .79$\pm$.27 (3) & \cellcolor{green!20}.95$\pm$.02 (1) & .93$\pm$.08 (2) & .94$\pm$.03 (2) & \cellcolor{green!20}.89$\pm$.05 (1) & .85$\pm$.15 (2) & .83$\pm$.20 (2) \\
& \textsc{DeepArch} & \cellcolor{green!20}1.0$\pm$.00 (1) & 1.0$\pm$.01 (2) & \cellcolor{green!20}1.0$\pm$.00 (1) & \cellcolor{green!20}1.0$\pm$.00 (1) & 1.0$\pm$.01 (2) & 1.0$\pm$.01 (2) & \cellcolor{green!20}1.0$\pm$.00 (1) & \cellcolor{green!20}1.0$\pm$.00 (1) & 1.0$\pm$.01 (2) \\
& \textsc{Jump3r} & \cellcolor{green!20}.27$\pm$.40 (1) & .04$\pm$.17 (2) & .04$\pm$.17 (2) & .15$\pm$.32 (2) & \cellcolor{green!20}.25$\pm$.40 (1) & .02$\pm$.02 (3) & \cellcolor{green!20}.10$\pm$.23 (1) & .03$\pm$.02 (2) & .04$\pm$.14 (2) \\
& \textsc{HSMGP} & \cellcolor{green!20}1.0$\pm$.00 (1) & \cellcolor{green!20}1.0$\pm$.00 (1) & .92$\pm$.23 (2) & \cellcolor{green!20}.98$\pm$.13 (1) & \cellcolor{green!20}.98$\pm$.13 (1) & \cellcolor{green!20}.98$\pm$.13 (1) & \cellcolor{green!20}.97$\pm$.14 (1) & \cellcolor{green!20}.95$\pm$.19 (1) & .92$\pm$.23 (2) \\


       \hline       
      \multirow{9}{*}{$50\%$}
& \textsc{7z} & \cellcolor{green!20}1.0$\pm$.00 (1) & \cellcolor{green!20}1.0$\pm$.00 (1) & \cellcolor{green!20}1.0$\pm$.00 (1) & \cellcolor{green!20}.64$\pm$.28 (1) & \cellcolor{green!20}.63$\pm$.29 (1) & \cellcolor{green!20}.67$\pm$.27 (1) & \cellcolor{green!20}.93$\pm$.18 (1) & .82$\pm$.28 (2) & .80$\pm$.34 (2) \\
& \textsc{Kanzi} & \cellcolor{green!20}.64$\pm$.14 (1) & .43$\pm$.22 (2) & .24$\pm$.24 (3) & \cellcolor{green!20}.64$\pm$.22 (1) & .46$\pm$.24 (2) & .42$\pm$.29 (2) & \cellcolor{green!20}.50$\pm$.15 (1) & .32$\pm$.12 (2) & .22$\pm$.17 (3) \\
& \textsc{ExaStencils} & \cellcolor{green!20}.99$\pm$.04 (1) & \cellcolor{green!20}.98$\pm$.04 (1) & .96$\pm$.15 (2) & \cellcolor{green!20}.88$\pm$.15 (1) & .63$\pm$.39 (3) & .74$\pm$.34 (2) & .99$\pm$.02 (2) & .98$\pm$.02 (3) & \cellcolor{green!20}.99$\pm$.00 (1) \\
& \textsc{Apache} & \cellcolor{green!20}.70$\pm$.03 (1) & .67$\pm$.04 (2) & .68$\pm$.04 (2) & \cellcolor{green!20}.34$\pm$.17 (1) & .24$\pm$.08 (3) & .26$\pm$.09 (2) & \cellcolor{green!20}.69$\pm$.07 (1) & .64$\pm$.10 (2) & .63$\pm$.08 (2) \\
& \textsc{SQLite} & \cellcolor{green!20}.74$\pm$.16 (1) & .60$\pm$.22 (2) & .55$\pm$.23 (3) & \cellcolor{green!20}.68$\pm$.19 (1) & .60$\pm$.15 (2) & .57$\pm$.15 (2) & \cellcolor{green!20}.80$\pm$.21 (1) & .62$\pm$.25 (2) & .56$\pm$.25 (3) \\
& \textsc{DConvert} & \cellcolor{green!20}.89$\pm$.04 (1) & .84$\pm$.17 (2) & \cellcolor{green!20}.88$\pm$.11 (1) & \cellcolor{green!20}.94$\pm$.02 (1) & .92$\pm$.11 (2) & .89$\pm$.15 (3) & \cellcolor{green!20}.93$\pm$.03 (1) & .92$\pm$.03 (2) & .83$\pm$.23 (3) \\
& \textsc{DeepArch} & \cellcolor{green!20}1.0$\pm$.00 (1) & 1.0$\pm$.01 (2) & \cellcolor{green!20}1.0$\pm$.00 (1) & \cellcolor{green!20}1.0$\pm$.00 (1) & \cellcolor{green!20}1.0$\pm$.00 (1) & \cellcolor{green!20}1.0$\pm$.00 (1) & \cellcolor{green!20}1.0$\pm$.00 (1) & \cellcolor{green!20}1.0$\pm$.00 (1) & 1.0$\pm$.01 (2) \\
& \textsc{Jump3r} & \cellcolor{green!20}.45$\pm$.33 (1) & .28$\pm$.22 (2) & .14$\pm$.17 (3) & \cellcolor{green!20}.42$\pm$.25 (1) & .30$\pm$.15 (2) & .23$\pm$.11 (3) & \cellcolor{green!20}.35$\pm$.24 (1) & .24$\pm$.11 (2) & .16$\pm$.10 (3) \\
& \textsc{HSMGP} & \cellcolor{green!20}1.0$\pm$.00 (1) & .99$\pm$.07 (2) & .96$\pm$.13 (3) & \cellcolor{green!20}1.0$\pm$.00 (1) & .97$\pm$.10 (3) & .99$\pm$.07 (2) & \cellcolor{green!20}.98$\pm$.11 (1) & \cellcolor{green!20}.96$\pm$.15 (1) & \cellcolor{green!20}.96$\pm$.15 (1) \\


       \hline       

            \multirow{9}{*}{$90\%$}
& \textsc{7z} & \cellcolor{green!20}.77$\pm$.31 (1) & \cellcolor{green!20}.74$\pm$.35 (1) & \cellcolor{green!20}.73$\pm$.34 (1) & \cellcolor{green!20}.81$\pm$.26 (1) & \cellcolor{green!20}.83$\pm$.30 (1) & \cellcolor{green!20}.77$\pm$.30 (1) & \cellcolor{green!20}.94$\pm$.15 (1) & .87$\pm$.24 (2) & .69$\pm$.29 (3) \\
& \textsc{Kanzi} & \cellcolor{green!20}.46$\pm$.19 (1) & .32$\pm$.20 (2) & .20$\pm$.18 (3) & \cellcolor{green!20}.30$\pm$.22 (1) & .21$\pm$.19 (2) & .09$\pm$.16 (3) & \cellcolor{green!20}.30$\pm$.28 (1) & .19$\pm$.23 (2) & .09$\pm$.11 (3) \\
& \textsc{ExaStencils} & \cellcolor{green!20}1.0$\pm$.02 (1) & \cellcolor{green!20}1.0$\pm$.02 (1) & .98$\pm$.04 (2) & \cellcolor{green!20}.91$\pm$.14 (1) & \cellcolor{green!20}.91$\pm$.15 (1) & \cellcolor{green!20}.92$\pm$.12 (1) & \cellcolor{green!20}.98$\pm$.04 (1) & .97$\pm$.06 (2) & .97$\pm$.06 (2) \\
& \textsc{Apache} & \cellcolor{green!20}.99$\pm$.00 (1) & .99$\pm$.01 (2) & .99$\pm$.01 (2) & \cellcolor{green!20}.80$\pm$.00 (1) & .80$\pm$.01 (2) & .80$\pm$.01 (2) & \cellcolor{green!20}.98$\pm$.00 (1) & .98$\pm$.01 (2) & .98$\pm$.01 (3) \\
& \textsc{SQLite} & \cellcolor{green!20}.63$\pm$.18 (1) & .51$\pm$.19 (2) & .46$\pm$.18 (3) & \cellcolor{green!20}.68$\pm$.17 (1) & .57$\pm$.14 (2) & .53$\pm$.11 (3) & \cellcolor{green!20}.78$\pm$.22 (1) & .50$\pm$.22 (3) & .60$\pm$.21 (2) \\
& \textsc{DConvert} & \cellcolor{green!20}.88$\pm$.04 (1) & .88$\pm$.10 (2) & .87$\pm$.10 (2) & \cellcolor{green!20}.94$\pm$.03 (1) & .91$\pm$.15 (2) & .89$\pm$.19 (2) & \cellcolor{green!20}.88$\pm$.13 (1) & \cellcolor{green!20}.86$\pm$.19 (1) & \cellcolor{green!20}.89$\pm$.14 (1) \\
& \textsc{DeepArch} & \cellcolor{green!20}1.0$\pm$.00 (1) & \cellcolor{green!20}1.0$\pm$.00 (1) & 1.0$\pm$.01 (2) & \cellcolor{green!20}1.0$\pm$.00 (1) & \cellcolor{green!20}1.0$\pm$.00 (1) & \cellcolor{green!20}1.0$\pm$.00 (1) & \cellcolor{green!20}1.0$\pm$.00 (1) & \cellcolor{green!20}1.0$\pm$.00 (1) & \cellcolor{green!20}1.0$\pm$.00 (1) \\
& \textsc{Jump3r} & \cellcolor{green!20}.32$\pm$.17 (1) & .25$\pm$.13 (2) & .20$\pm$.15 (3) & \cellcolor{green!20}.27$\pm$.38 (1) & .05$\pm$.01 (2) & .04$\pm$.01 (3) & \cellcolor{green!20}.09$\pm$.20 (1) & .05$\pm$.13 (2) & .02$\pm$.01 (3) \\
& \textsc{HSMGP} & \cellcolor{green!20}1.0$\pm$.00 (1) & .88$\pm$.27 (2) & .93$\pm$.22 (2) & \cellcolor{green!20}1.0$\pm$.00 (1) & .89$\pm$.27 (2) & .92$\pm$.24 (2) & \cellcolor{green!20}.98$\pm$.13 (1) & .93$\pm$.21 (2) & .93$\pm$.22 (2) \\

      \hline
\multicolumn{2}{c}{Average $p_t$ score/rank} & \multicolumn{1}{l}{\cellcolor{green!20}.62/1.09} & \multicolumn{1}{l}{.52/1.87} & \multicolumn{1}{l|}{.55/1.94} & \multicolumn{1}{l}{\cellcolor{green!20}.57/1.20} & \multicolumn{1}{l}{.48/1.83} & \multicolumn{1}{l|}{.53/1.85} & \multicolumn{1}{l}{\cellcolor{green!20}.56/1.13} & \multicolumn{1}{l}{.47/1.81} & \multicolumn{1}{l}{.51/1.87} \\

      \bottomrule
    \end{tabular}
  \end{adjustbox}
  \vspace{-0.3cm}
\end{table}

%


\subsubsection{Method}

For \textbf{RQ1}, we evaluate \tuner~against the two variants of \texttt{GA} on all nine systems and 18 target performance requirements, leading to 162 cases tested by Scott-Knott ESD. Since the configuration evolution part in \tuner~is essentially a \texttt{GA}, this helps to examine the necessity of co-evolution.


\subsubsection{Results}

As shown in Table~\ref{tab:rq1}, we see that \tuner~significantly outperforms the others in general, regardless of the systems and requirements. Within the 162 cases, \tuner~has the best Scott-Knott ESD rank in $90\%$ of them ($50\%$ sole best) against the $35\%$ ($3\%$ sole best) and $35\%$ ($5\%$ sole best) for \texttt{GA}$_p$ and \texttt{GA}$_r$, respectively. \tuner~also has the best overall ranks and satisfaction with up to $1.78\times$ overall improvement. Interestingly, we see that \texttt{GA}$_p$ is generally better than \texttt{GA}$_r$ on easier requirements ($d\geq5\%$), but worse on harder ones. This makes sense, as the requirement is useful when it is reasonable and hence \texttt{GA}$_p$ is advantaged. However, the requirement can most commonly mislead the tuning when it becomes too strict, while \texttt{GA}$_r$ is not affected. \tuner, in contrast, performs the best in both situations, thanks to the co-evolution. 

\begin{figure}[t!]
\centering
\begin{tikzpicture}
    \begin{axis}[
      axis x line  = bottom,
            axis y line  = left  ,
        width=10cm, height=4cm, 
        xlabel={Budget},
          ylabel={Avg. ${p}_t$ satisfaction score},
        xmin=0, xmax=300,
        xtick={20,60,100,140,180,220,260,300},
        ymin=0, ymax=0.6,
        ytick={-0.2, 0, 0.2, 0.4, 0.6, 0.8, 1.0},
        grid=minor,
        legend style={
          draw=none, fill=none,
          at={(1.0,0.3)}, anchor=south east,
          legend cell align=left,
          legend columns=3
        }
    ]


\addplot[
    red,
    mark=triangle*,mark options={fill=red!50,draw=black},
    thick
] coordinates {
(0,0)
(20,0.26853721)
(30,0.32950946)
(40,0.36682536)
(50,0.38968223)
(60,0.40593088)
(70,0.41837047)
(80,0.42917769)
(90,0.43781918)
(100,0.44495712)
(110,0.44976665)
(120,0.45376652)
(130,0.45676914)
(140,0.45926741)
(150,0.46131584)
(160,0.46290411)
(170,0.46394326)
(180,0.46490170)
(190,0.46628323)
(200,0.46755899)
(210,0.46885283)
(220,0.47035093)
(230,0.47113508)
(240,0.47174835)
(250,0.47303315)
(260,0.47420703)
(270,0.47488600)
(280,0.47588090)
(290,0.47619247)
(300,0.47657766)
};\addlegendentry{\texttt{GA}$_{p}$}

\addplot[
    teal,
     mark=square*,mark options={fill=teal!50,draw=black},
    thick
] coordinates {
(0,0)
(20,0.25871913)
(30,0.35135955)
(40,0.40507966)
(50,0.43678886)
(60,0.45871155)
(70,0.47836601)
(80,0.49332091)
(90,0.50536894)
(100,0.51349574)
(110,0.51946758)
(120,0.52334904)
(130,0.52603452)
(140,0.52701931)
(150,0.52784646)
(160,0.52806097)
(170,0.52832944)
(180,0.52835670)
(190,0.52836012)
(200,0.52836012)
(210,0.52836181)
(220,0.52836181)
(230,0.52836181)
(240,0.52836181)
(250,0.52836181)
(260,0.52836181)
(270,0.52836181)
(280,0.52836181)
(290,0.52836181)
(300,0.52836181)
};\addlegendentry{\texttt{GA}$_{r}$}

   \addplot[
    blue,
    mark=*,mark options={fill=blue!50,draw=black},
    thick
] coordinates {
(0,0)
(20,0.21737327)
(30,0.34115952)
(40,0.41724135)
(50,0.46907641)
(60,0.49794897)
(70,0.51668949)
(80,0.53178391)
(90,0.54374004)
(100,0.55332311)
(110,0.56094469)
(120,0.56655645)
(130,0.57016500)
(140,0.57392688)
(150,0.57610281)
(160,0.57820866)
(170,0.57987429)
(180,0.58078162)
(190,0.58138777)
(200,0.58165077)
(210,0.58244060)
(220,0.58290687)
(230,0.58303468)
(240,0.58319771)
(250,0.58331187)
(260,0.58331187)
(270,0.58341192)
(280,0.58343312)
(290,0.58348804)
(300,0.58348804)
};\addlegendentry{\tuner}

    \end{axis}
\end{tikzpicture}
\caption{Trajectories of \tuner~and \texttt{GA} variants on different budgets over all cases/runs.}
\label{fig:rq1}
\vspace{-0.3cm}
\end{figure}
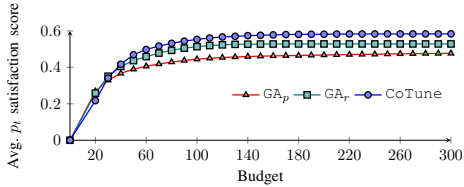

The same can also be observed in Figure~\ref{fig:rq1}, where the tuning trajectory of \tuner~is generally better/steeper than the other two. This means that even if we have a different budget, \tuner~would have still maintained its superiority.



\begin{rbox}
   \textit{\textbf{RQ1:} \tuner, with its co-evolution, significantly outperforms the single-evolution \texttt{GA} with and without requirement, having the best rank on $90\%$ of the cases (vs. $35\%$), boosting the overall satisfaction up to $1.78\times$ with better efficiency.}
\end{rbox}


\subsection{Improvements over State-of-the-art Tuners}

\subsubsection{Method}

We compare \tuner~with the other state-of-the-art tuners under two of their variants as part of \textbf{RQ2}. We follow the same procedure for \textbf{RQ1} except that we aggregate results for the three requirement types due to space constraint. 





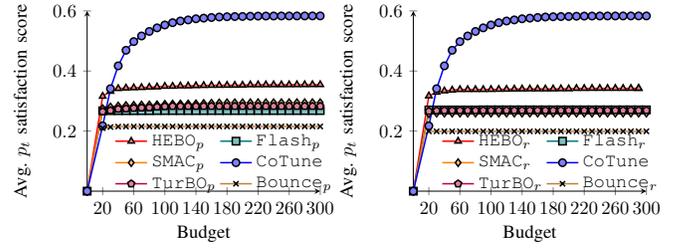
\begin{figure}[t!]
\centering
\subfloat[\tuner~vs. w/ requirement]{
\begin{tikzpicture}
    \begin{axis}[
      axis x line  = bottom,
            axis y line  = left  ,
        width=6cm, height=5cm, 
          xlabel={Budget},
        ylabel={Avg. ${p}_t$ satisfaction score},
        xmin=0, xmax=300,
        xtick={20,60,100,140,180,220,260,300},
        ymin=0, ymax=0.6,
        ytick={0.2, 0.4, 0.6, 0.8, 1.0},
        grid=minor,
        legend style={
          draw=none, fill=none,
          at={(1.09,-0.04)}, anchor=south east,
          legend cell align=left, legend columns=2
        }
    ]

\addplot[
    red,
    mark=triangle*,mark options={fill=red!50,draw=black},
    thick
] coordinates {
(0,0)
(20,0.31596188)
(30,0.33239086)
(40,0.34168911)
(50,0.34335625)
(60,0.34419035)
(70,0.34502547)
(80,0.34623539)
(90,0.34706171)
(100,0.34926775)
(110,0.34932576)
(120,0.35036776)
(130,0.35040474)
(140,0.35120208)
(150,0.35189479)
(160,0.35232089)
(170,0.35241424)
(180,0.35257489)
(190,0.35258468)
(200,0.35296549)
(210,0.35404322)
(220,0.35405100)
(230,0.35419821)
(240,0.35420860)
(250,0.35422312)
(260,0.35426585)
(270,0.35439083)
(280,0.35446880)
(290,0.35463766)
(300,0.35467687)
};\addlegendentry{\texttt{HEBO}$_p$}

\addplot[
    teal,
     mark=square*,mark options={fill=teal!50,draw=black},
    thick
] coordinates {
(0,0)
(20,0.26771882)
(30,0.26803079)
(40,0.26803079)
(50,0.26803079)
(60,0.26803123)
(70,0.26803697)
(80,0.26803697)
(90,0.26833942)
(100,0.26833942)
(110,0.26833942)
(120,0.26833942)
(130,0.26833942)
(140,0.26833942)
(150,0.26833942)
(160,0.26838971)
(170,0.26838971)
(180,0.26838971)
(190,0.26838971)
(200,0.26838971)
(210,0.26838971)
(220,0.26838971)
(230,0.26838971)
(240,0.26838971)
(250,0.26838971)
(260,0.26838971)
(270,0.26838971)
(280,0.26838971)
(290,0.26838971)
(300,0.26838971)
};\addlegendentry{\texttt{Flash}$_p$}

   \addplot[
    orange,
    mark=diamond*,mark options={fill=orange!50,draw=black},
    thick
] coordinates {
(0,0)
(20,0.27472194)
(30,0.28037670)
(40,0.28365984)
(50,0.28515808)
(60,0.28582335)
(70,0.28687812)
(80,0.28703829)
(90,0.28780261)
(100,0.28928804)
(110,0.28997270)
(120,0.29031850)
(130,0.29037438)
(140,0.29083284)
(150,0.29206127)
(160,0.29297352)
(170,0.29386327)
(180,0.29437543)
(190,0.29437696)
(200,0.29439061)
(210,0.29449912)
(220,0.29451908)
(230,0.29453866)
(240,0.29453866)
(250,0.29458119)
(260,0.29459095)
(270,0.29459095)
(280,0.29459095)
(290,0.29469629)
(300,0.29469629)
};\addlegendentry{\texttt{SMAC}$_p$}

   \addplot[
    blue,
    mark=*,mark options={fill=blue!50,draw=black},
    thick
] coordinates {
(0,0)
(20,0.21737327)
(30,0.34115952)
(40,0.41724135)
(50,0.46907641)
(60,0.49794897)
(70,0.51668949)
(80,0.53178391)
(90,0.54374004)
(100,0.55332311)
(110,0.56094469)
(120,0.56655645)
(130,0.57016500)
(140,0.57392688)
(150,0.57610281)
(160,0.57820866)
(170,0.57987429)
(180,0.58078162)
(190,0.58138777)
(200,0.58165077)
(210,0.58244060)
(220,0.58290687)
(230,0.58303468)
(240,0.58319771)
(250,0.58331187)
(260,0.58331187)
(270,0.58341192)
(280,0.58343312)
(290,0.58348804)
(300,0.58348804)
};\addlegendentry{\texttt{CoTune}}

\addplot[
    purple,
    mark=pentagon*,mark options={fill=purple!50,draw=black},
    thick
] coordinates {
(20,0.26485070)
(30,0.26859528)
(40,0.27359528)
(50,0.27459528)
(60,0.27653308)
(70,0.2776256)
(80,0.27827949)
(90,0.28057935)
(100,0.28189329)
(110,0.28277775)
(120,0.28277775)
(130,0.28277775)
(140,0.28277775)
(150,0.28277775)
(160,0.28277775)
(170,0.28277775)
(180,0.28277775)
(190,0.28277775)
(200,0.28277775)
(210,0.28277775)
(220,0.28277775)
(230,0.28277775)
(240,0.28277775)
(250,0.28277775)
(260,0.28277775)
(270,0.28277775)
(280,0.28277775)
(290,0.28277775)
(300,0.28277775)
};\addlegendentry{\texttt{TurBO}$_p$}

\addplot[
    brown,
    mark=x,mark options={fill=brown,draw=black},
    thick
] coordinates {
(20,0.20979624)
(30,0.21348528)
(40,0.21492167)
(50,0.21517678)
(60,0.21521179)
(70,0.21530373)
(80,0.21541130)
(90,0.21541248)
(100,0.21541248)
(110,0.21541248)
(120,0.21541248)
(130,0.21541248)
(140,0.21541248)
(150,0.21541248)
(160,0.21541248)
(170,0.21541248)
(180,0.21541248)
(190,0.21541248)
(200,0.21541248)
(210,0.21541248)
(220,0.21541248)
(230,0.21541248)
(240,0.21541248)
(250,0.21541248)
(260,0.21541248)
(270,0.21541248)
(280,0.21541248)
(290,0.21541248)
(300,0.21541248)
};\addlegendentry{\texttt{Bounce}$_p$}

    \end{axis}
\end{tikzpicture}
}
~\hspace{-0.5cm}
\subfloat[\tuner~vs. w/o requirement]{
\begin{tikzpicture}
    \begin{axis}[
      axis x line  = bottom,
            axis y line  = left  ,
        width=6cm, height=5cm, 
          xlabel={Budget},
          ylabel={Avg. ${p}_t$ satisfaction score},
        xmin=0, xmax=300,
        xtick={20,60,100,140,180,220,260,300},
        ymin=0, ymax=0.6,
        ytick={0.2, 0.4, 0.6, 0.8, 1.0},
        grid=minor,
        legend style={
          draw=none, fill=none,
          at={(1.09,-0.04)}, anchor=south east,
          legend cell align=left, legend columns=2
        }
    ]


\addplot[
    red,
    mark=triangle*,mark options={fill=red!50,draw=black},
    thick
] coordinates {
(0,0)
(20,0.31719607)
(30,0.32943084)
(40,0.33391528)
(50,0.33651935)
(60,0.33834977)
(70,0.33837370)
(80,0.33839508)
(90,0.33841497)
(100,0.33861382)
(110,0.33908363)
(120,0.33938413)
(130,0.34004216)
(140,0.34011443)
(150,0.34039574)
(160,0.34039815)
(170,0.34039950)
(180,0.34043266)
(190,0.34077047)
(200,0.34077047)
(210,0.34206013)
(220,0.34206013)
(230,0.34226103)
(240,0.34226103)
(250,0.34226270)
(260,0.34226473)
(270,0.34226473)
(280,0.34226548)
(290,0.34228948)
(290,0.34228948)
};\addlegendentry{\texttt{HEBO}$_r$}

\addplot[
    teal,
     mark=square*,mark options={fill=teal!50,draw=black},
    thick
] coordinates {
(0,0)
(20,0.27032850)
(30,0.27032850)
(40,0.27032850)
(50,0.27032850)
(60,0.27032850)
(70,0.27032850)
(80,0.27032850)
(90,0.27075726)
(100,0.27075726)
(110,0.27075726)
(120,0.27075726)
(130,0.27075726)
(140,0.27075726)
(150,0.27075726)
(160,0.27075726)
(170,0.27075726)
(180,0.27075726)
(190,0.27075726)
(200,0.27075726)
(210,0.27075726)
(220,0.27075726)
(230,0.27075726)
(240,0.27075726)
(250,0.27075726)
(260,0.27075726)
(270,0.27075726)
(280,0.27075726)
(290,0.27075726)
(300,0.27075726)
};\addlegendentry{\texttt{Flash}$_r$}

   \addplot[
    orange,
    mark=diamond*,mark options={fill=orange!50,draw=black},
    thick
] coordinates {
(0,0)
(20,0.25777625)
(30,0.25983593)
(40,0.25983593)
(50,0.25983593)
(60,0.25983593)
(70,0.25983593)
(80,0.25983593)
(90,0.25983593)
(100,0.25983593)
(110,0.25983593)
(120,0.25983593)
(130,0.25983593)
(140,0.25983593)
(150,0.25983593)
(160,0.25983593)
(170,0.25983593)
(180,0.25983593)
(190,0.25983593)
(200,0.25983593)
(210,0.25983593)
(220,0.25983593)
(230,0.25983593)
(240,0.25983593)
(250,0.25983593)
(260,0.25983593)
(270,0.25983593)
(280,0.25983593)
(290,0.25983593)
(300,0.25983593)
};\addlegendentry{\texttt{SMAC}$_r$}

   \addplot[
    blue,
    mark=*,mark options={fill=blue!50,draw=black},
    thick
] coordinates {
(0,0)
(20,0.21737327)
(30,0.34115952)
(40,0.41724135)
(50,0.46907641)
(60,0.49794897)
(70,0.51668949)
(80,0.53178391)
(90,0.54374004)
(100,0.55332311)
(110,0.56094469)
(120,0.56655645)
(130,0.57016500)
(140,0.57392688)
(150,0.57610281)
(160,0.57820866)
(170,0.57987429)
(180,0.58078162)
(190,0.58138777)
(200,0.58165077)
(210,0.58244060)
(220,0.58290687)
(230,0.58303468)
(240,0.58319771)
(250,0.58331187)
(260,0.58331187)
(270,0.58341192)
(280,0.58343312)
(290,0.58348804)
(300,0.58348804)
};\addlegendentry{\texttt{CoTune}}

\addplot[
    purple,
    mark=pentagon*,mark options={fill=purple!50,draw=black},
    thick
] coordinates {
(20,0.26614429)
(30,0.26794944)
(40,0.26850174)
(50,0.26854724)
(60,0.26859154)
(70,0.26859154)
(80,0.26862347)
(90,0.26862347)
(100,0.26864347)
(110,0.26868347)
(120,0.26874347)
(130,0.26883242)
(140,0.26883242)
(150,0.26883242)
(160,0.26883242)
(170,0.26883242)
(180,0.26883242)
(190,0.26883242)
(200,0.26883242)
(210,0.26883242)
(220,0.26883242)
(230,0.26883242)
(240,0.26883242)
(250,0.26883242)
(260,0.26883242)
(270,0.26883242)
(280,0.26883242)
(290,0.26883242)
(300,0.26883242)
};\addlegendentry{\texttt{TurBO}$_r$}

\addplot[
    brown,
    mark=x,mark options={fill=brown,draw=black},
    thick
] coordinates {
(20,0.19912566)
(30,0.19917129)
(40,0.19918930)
(50,0.19918930)
(60,0.19918930)
(70,0.19918930)
(80,0.19918930)
(90,0.19918930)
(100,0.19918930)
(110,0.19918930)
(120,0.19918930)
(130,0.19918930)
(140,0.19918930)
(150,0.19918930)
(160,0.19918930)
(170,0.19918930)
(180,0.19918930)
(190,0.19918930)
(200,0.19918930)
(210,0.19918930)
(220,0.19918930)
(230,0.19918930)
(240,0.19918930)
(250,0.19918930)
(260,0.19918930)
(270,0.19918930)
(280,0.19918930)
(290,0.19918930)
(300,0.19918930)
};\addlegendentry{\texttt{Bounce}$_r$}

    \end{axis}
\end{tikzpicture}
}

\caption{Trajectories of \tuner~and state-of-the-art tuners (w/ and w/o requirement) on different budgets over all cases/runs.}
\label{fig:rq2}
\vspace{-0.3cm}
\end{figure}

\begin{table*}[t!]
  \setlength{\tabcolsep}{3mm}
  \caption{\revision{Comparing \tuner~with state-of-the-art tuners under 300 budgets/30 runs by averaging all three types of requirements $p_{t,1}$, $p_{t,2}$, and $p_{t,3}$. \textcolor{red}{\ding{55}} denotes failed to complete in a reasonable time. The format follows Table~\ref{tab:rq1}. More detailed data can be found at: \textcolor{blue}{\texttt{\url{https://github.com/ideas-labo/CoTune/blob/main/rqSupplementary/RQ2.pdf}}}.}}

  \label{tab:rq2}

  \begin{adjustbox}{max width=\textwidth,center}
    \centering
    \scriptsize
    \begin{tabular}{lllllllllllll}
      \toprule
       \textbf{${d\%}$}&
       \textbf{System}&\tuner&\texttt{HEBO}$_p$&\texttt{HEBO}$_r$&\texttt{Flash}$_p$&\texttt{Flash}$_r$&\texttt{SMAC}$_p$&\texttt{SMAC}$_r$&\texttt{TurBO}$_p$&\texttt{TurBO}$_r$ 
       &\texttt{Bounce}$_p$&\texttt{Bounce}$_r$ \\
      \midrule

\multirow{9}{*}{$0.1\%$}
& \textsc{7z} & \cellcolor{green!20}.28$\pm$.35 (1.00) & .00$\pm$.00 (2.00) & .00$\pm$.00 (2.00) & .00$\pm$.00 (2.00) & .00$\pm$.00 (2.00) & .00$\pm$.00 (2.00) & .00$\pm$.00 (2.00) & \cellcolor{green!20}.00$\pm$.01 (1.00) & .00$\pm$.00 (2.00) & .00$\pm$.00 (2.00) & .00$\pm$.00 (2.00) \\
& \textsc{Kanzi} & \cellcolor{green!20}.00$\pm$.02 (1.00) & .00$\pm$.00 (1.33) & .00$\pm$.00 (1.33) & \textcolor{red}{\ding{55}} & \textcolor{red}{\ding{55}} & .00$\pm$.00 (1.33) & .00$\pm$.00 (1.33) & .00$\pm$.00 (1.33) & .00$\pm$.00 (1.33) & .00$\pm$.00 (1.33) & .00$\pm$.00 (1.33) \\
& \textsc{ExaStencils} & \cellcolor{green!20}.67$\pm$.47 (1.00) & .00$\pm$.00 (2.00) & .00$\pm$.00 (2.00) & .00$\pm$.00 (2.00) & .00$\pm$.00 (2.00) & .00$\pm$.00 (2.00) & .00$\pm$.00 (2.00) & .00$\pm$.00 (2.00) & .00$\pm$.00 (2.00) & .00$\pm$.00 (2.00) & .00$\pm$.00 (2.00) \\
& \textsc{Apache} & \cellcolor{green!20}.00$\pm$.00 (1.00) & .00$\pm$.00 (1.33) & .00$\pm$.00 (1.33) & .00$\pm$.00 (1.33) & .00$\pm$.00 (1.33) & .00$\pm$.00 (1.33) & .00$\pm$.00 (1.33) & .00$\pm$.00 (1.33) & .00$\pm$.00 (1.33) & .00$\pm$.00 (1.33) & .00$\pm$.00 (1.33) \\
& \textsc{SQLite} & \cellcolor{green!20}.01$\pm$.07 (1.00) & .00$\pm$.00 (2.00) & .00$\pm$.00 (2.00) & \textcolor{red}{\ding{55}} & \textcolor{red}{\ding{55}} & .00$\pm$.00 (2.00) & .00$\pm$.00 (2.00) & .00$\pm$.00 (2.00) & .00$\pm$.00 (2.00) & .00$\pm$.00 (2.00) & .00$\pm$.00 (2.00) \\
& \textsc{DConvert} & \cellcolor{green!20}.01$\pm$.06 (1.00) & .00$\pm$.00 (1.33) & .00$\pm$.00 (1.33) & .00$\pm$.00 (1.33) & .00$\pm$.00 (1.33) & .00$\pm$.00 (1.33) & .00$\pm$.00 (1.33) & .00$\pm$.00 (1.33) & .00$\pm$.00 (1.33) & .00$\pm$.00 (1.33) & .00$\pm$.00 (1.33) \\
& \textsc{DeepArch} & \cellcolor{green!20}.70$\pm$.42 (1.00) & .10$\pm$.23 (2.00) & .00$\pm$.00 (3.67) & .00$\pm$.00 (3.67) & .00$\pm$.00 (3.67) & .10$\pm$.21 (2.00) & .00$\pm$.00 (3.67) & .00$\pm$.00 (3.67) & .00$\pm$.00 (3.67) & .01$\pm$.03 (3.00) & .00$\pm$.00 (3.67) \\
& \textsc{Jump3r} & \cellcolor{green!20}.00$\pm$.00 (1.00) & .00$\pm$.00 (1.67) & .00$\pm$.00 (1.67) & \textcolor{red}{\ding{55}} & \textcolor{red}{\ding{55}} & .00$\pm$.00 (1.67) & .00$\pm$.00 (1.67) & .00$\pm$.00 (1.67) & .00$\pm$.00 (1.67) & .00$\pm$.00 (1.67) & .00$\pm$.00 (1.67) \\
& \textsc{HSMGP} & \cellcolor{green!20}.81$\pm$.35 (1.00) & .00$\pm$.00 (2.00) & .00$\pm$.00 (2.33) & .00$\pm$.00 (2.33) & .00$\pm$.00 (2.33) & .00$\pm$.01 (1.67) & .00$\pm$.00 (2.33) & .00$\pm$.00 (2.33) & .00$\pm$.00 (2.33) & .00$\pm$.00 (2.33) & .00$\pm$.00 (2.33) \\

\hline
\multirow{9}{*}{$1\%$}
& \textsc{7z} & \cellcolor{green!20}.35$\pm$.29 (1.00) & .01$\pm$.03 (3.33) & .01$\pm$.03 (3.33) & .01$\pm$.03 (3.33) & .01$\pm$.03 (3.33) & .01$\pm$.03 (3.33) & .01$\pm$.03 (3.33) & .05$\pm$.13 (2.00) & .01$\pm$.03 (3.33) & .01$\pm$.08 (2.67) & .01$\pm$.05 (3.33) \\
& \textsc{Kanzi} & \cellcolor{green!20}.07$\pm$.22 (1.00) & .03$\pm$.14 (1.67) & .03$\pm$.14 (1.67) & \textcolor{red}{\ding{55}} & \textcolor{red}{\ding{55}} & .03$\pm$.14 (1.67) & .03$\pm$.14 (1.67) & .03$\pm$.14 (1.67) & .03$\pm$.14 (1.67) & .00$\pm$.00 (2.67) & .00$\pm$.00 (2.67) \\
& \textsc{ExaStencils} & \cellcolor{green!20}.91$\pm$.11 (1.00) & .08$\pm$.17 (3.33) & .27$\pm$.33 (2.00) & .03$\pm$.12 (4.00) & .03$\pm$.12 (4.00) & .11$\pm$.20 (3.00) & .03$\pm$.11 (4.00) & .01$\pm$.03 (4.67) & .04$\pm$.12 (4.00) & .01$\pm$.03 (5.33) & .00$\pm$.00 (6.33) \\
& \textsc{Apache} & .02$\pm$.08 (4.33) & .33$\pm$.35 (1.33) & .38$\pm$.39 (1.33) & .03$\pm$.10 (4.00) & .02$\pm$.10 (4.33) & .15$\pm$.26 (2.67) & .05$\pm$.18 (3.67) & .28$\pm$.33 (1.67) & .02$\pm$.10 (4.33) & .29$\pm$.33 (1.67) & .24$\pm$.32 (1.67) \\
& \textsc{SQLite} & \cellcolor{green!20}.17$\pm$.27 (1.00) & .04$\pm$.13 (2.33) & .05$\pm$.15 (2.00) & \textcolor{red}{\ding{55}} & \textcolor{red}{\ding{55}} & .04$\pm$.14 (2.33) & .05$\pm$.15 (2.00) & .05$\pm$.16 (2.00) & .05$\pm$.15 (2.00) & .00$\pm$.00 (3.33) & .00$\pm$.00 (3.33) \\
& \textsc{DConvert} & \cellcolor{green!20}.30$\pm$.13 (1.00) & .03$\pm$.07 (2.00) & .03$\pm$.07 (2.00) & .03$\pm$.07 (2.00) & .03$\pm$.07 (2.00) & .03$\pm$.07 (2.00) & .03$\pm$.07 (2.00) & .03$\pm$.07 (2.00) & .03$\pm$.07 (2.00) & .02$\pm$.06 (2.00) & .02$\pm$.06 (2.00) \\
& \textsc{DeepArch} & \cellcolor{green!20}.89$\pm$.20 (1.00) & .57$\pm$.33 (2.33) & .22$\pm$.07 (4.00) & .00$\pm$.00 (6.67) & .00$\pm$.00 (6.67) & .48$\pm$.36 (2.67) & .00$\pm$.00 (6.67) & .00$\pm$.00 (6.67) & .00$\pm$.00 (6.67) & .08$\pm$.22 (4.67) & .00$\pm$.01 (5.67) \\
& \textsc{Jump3r} & \cellcolor{green!20}.07$\pm$.18 (1.00) & .02$\pm$.12 (1.67) & .02$\pm$.12 (1.67) & \textcolor{red}{\ding{55}} & \textcolor{red}{\ding{55}} & .02$\pm$.12 (1.67) & .02$\pm$.12 (1.67) & .02$\pm$.12 (1.67) & .02$\pm$.12 (1.67) & .00$\pm$.00 (2.67) & .00$\pm$.00 (2.67) \\
& \textsc{HSMGP} & \cellcolor{green!20}.96$\pm$.15 (1.00) & .07$\pm$.20 (2.00) & .06$\pm$.20 (2.00) & .06$\pm$.20 (2.00) & .06$\pm$.20 (2.00) & .06$\pm$.20 (2.00) & .06$\pm$.20 (2.00) & .06$\pm$.20 (2.00) & .06$\pm$.20 (2.00) & .03$\pm$.15 (2.33) & .00$\pm$.00 (3.33) \\

\hline
\multirow{9}{*}{$5\%$}
& \textsc{7z} & \cellcolor{green!20}.64$\pm$.32 (1.00) & .18$\pm$.28 (3.33) & .19$\pm$.29 (3.33) & .19$\pm$.29 (3.33) & .19$\pm$.29 (3.33) & .33$\pm$.33 (2.00) & .18$\pm$.29 (3.33) & .25$\pm$.35 (2.67) & .19$\pm$.29 (3.33) & .03$\pm$.15 (4.67) & .03$\pm$.07 (4.33) \\
& \textsc{Kanzi} & .22$\pm$.30 (1.33) & .20$\pm$.28 (1.33) & .21$\pm$.29 (1.33) & \textcolor{red}{\ding{55}} & \textcolor{red}{\ding{55}} & .21$\pm$.29 (1.33) & .21$\pm$.29 (1.33) & .22$\pm$.29 (1.33) & .21$\pm$.29 (1.33) & .09$\pm$.16 (2.33) & .09$\pm$.16 (2.33) \\
& \textsc{ExaStencils} & \cellcolor{green!20}.94$\pm$.11 (1.00) & .26$\pm$.23 (2.67) & .33$\pm$.36 (2.00) & .07$\pm$.17 (5.33) & .07$\pm$.17 (5.33) & .21$\pm$.25 (3.00) & .06$\pm$.17 (5.33) & .13$\pm$.15 (4.00) & .11$\pm$.19 (4.33) & .02$\pm$.07 (6.33) & .00$\pm$.01 (7.33) \\
& \textsc{Apache} & .23$\pm$.12 (3.67) & \cellcolor{green!20}.48$\pm$.27 (1.00) & .26$\pm$.26 (3.00) & .06$\pm$.14 (5.00) & .06$\pm$.14 (5.00) & .22$\pm$.21 (3.67) & .08$\pm$.18 (5.00) & .42$\pm$.22 (2.00) & .07$\pm$.14 (5.00) & .42$\pm$.22 (1.67) & .39$\pm$.21 (2.00) \\
& \textsc{SQLite} & \cellcolor{green!20}.29$\pm$.29 (1.00) & .13$\pm$.20 (2.00) & .12$\pm$.20 (2.00) & \textcolor{red}{\ding{55}} & \textcolor{red}{\ding{55}} & .14$\pm$.20 (2.00) & .14$\pm$.20 (2.00) & .14$\pm$.20 (2.00) & .14$\pm$.20 (2.00) & .00$\pm$.00 (3.00) & .00$\pm$.00 (3.00) \\
& \textsc{DConvert} & \cellcolor{green!20}.67$\pm$.15 (1.00) & .19$\pm$.27 (2.00) & .17$\pm$.25 (2.00) & .18$\pm$.25 (2.00) & .18$\pm$.25 (2.00) & .18$\pm$.25 (2.00) & .18$\pm$.25 (2.00) & .18$\pm$.25 (2.00) & .17$\pm$.25 (2.00) & .07$\pm$.17 (3.00) & .07$\pm$.17 (3.00) \\
& \textsc{DeepArch} & \cellcolor{green!20}.99$\pm$.01 (1.00) & .93$\pm$.03 (2.00) & .92$\pm$.03 (3.00) & .15$\pm$.33 (6.00) & .15$\pm$.33 (6.00) & .75$\pm$.33 (4.00) & .15$\pm$.33 (6.00) & .15$\pm$.33 (6.00) & .15$\pm$.33 (6.00) & .30$\pm$.41 (5.00) & .05$\pm$.18 (7.00) \\
& \textsc{Jump3r} & \cellcolor{green!20}.14$\pm$.28 (1.00) & .06$\pm$.21 (1.67) & .06$\pm$.21 (1.67) & \textcolor{red}{\ding{55}} & \textcolor{red}{\ding{55}} & .06$\pm$.21 (1.67) & .06$\pm$.21 (1.67) & .06$\pm$.21 (1.67) & .06$\pm$.21 (1.67) & .03$\pm$.13 (2.00) & .03$\pm$.13 (2.00) \\
& \textsc{HSMGP} & \cellcolor{green!20}.99$\pm$.06 (1.00) & .29$\pm$.34 (2.00) & .23$\pm$.32 (2.67) & .21$\pm$.32 (3.00) & .20$\pm$.31 (3.33) & .29$\pm$.33 (2.00) & .20$\pm$.31 (3.33) & .20$\pm$.32 (3.00) & .21$\pm$.32 (3.00) & .17$\pm$.30 (3.67) & .12$\pm$.25 (4.00) \\

\hline
\multirow{9}{*}{$20\%$}
& \textsc{7z} & \cellcolor{green!20}.81$\pm$.17 (1.00) & .34$\pm$.31 (3.67) & .39$\pm$.31 (3.00) & .40$\pm$.30 (3.00) & .38$\pm$.31 (3.00) & .48$\pm$.26 (2.00) & .38$\pm$.31 (3.00) & .40$\pm$.35 (2.67) & .40$\pm$.31 (3.00) & .24$\pm$.31 (4.67) & .19$\pm$.21 (5.00) \\
& \textsc{Kanzi} & \cellcolor{green!20}.58$\pm$.24 (1.00) & .52$\pm$.25 (1.33) & .52$\pm$.25 (1.33) & \textcolor{red}{\ding{55}} & \textcolor{red}{\ding{55}} & .52$\pm$.26 (1.33) & .52$\pm$.25 (1.33) & .52$\pm$.25 (1.33) & .52$\pm$.25 (1.33) & .29$\pm$.29 (2.33) & .28$\pm$.29 (2.33) \\
& \textsc{ExaStencils} & \cellcolor{green!20}.98$\pm$.05 (1.00) & .61$\pm$.16 (3.00) & .70$\pm$.18 (2.00) & .25$\pm$.22 (6.00) & .24$\pm$.22 (6.00) & .30$\pm$.21 (5.00) & .22$\pm$.21 (6.00) & .35$\pm$.22 (4.00) & .28$\pm$.24 (5.00) & .07$\pm$.14 (7.00) & .03$\pm$.08 (8.00) \\
& \textsc{Apache} & .32$\pm$.06 (3.00) & .39$\pm$.19 (2.00) & .27$\pm$.09 (4.00) & .16$\pm$.14 (5.33) & .15$\pm$.14 (5.33) & .21$\pm$.15 (5.00) & .17$\pm$.18 (5.33) & \cellcolor{green!20}.45$\pm$.17 (1.00) & .17$\pm$.14 (5.33) & \cellcolor{green!20}.44$\pm$.16 (1.00) & \cellcolor{green!20}.43$\pm$.15 (1.00) \\
& \textsc{SQLite} & \cellcolor{green!20}.54$\pm$.20 (1.00) & .37$\pm$.18 (2.00) & .37$\pm$.18 (2.00) & \textcolor{red}{\ding{55}} & \textcolor{red}{\ding{55}} & .37$\pm$.19 (2.00) & .37$\pm$.19 (2.00) & .37$\pm$.19 (2.00) & .37$\pm$.19 (2.00) & .08$\pm$.13 (3.00) & .08$\pm$.13 (3.00) \\
& \textsc{DConvert} & \cellcolor{green!20}.92$\pm$.03 (1.00) & .49$\pm$.30 (2.00) & .33$\pm$.23 (3.00) & .34$\pm$.23 (3.00) & .31$\pm$.24 (3.00) & .34$\pm$.23 (3.00) & .31$\pm$.24 (3.00) & .34$\pm$.23 (3.00) & .32$\pm$.24 (3.00) & .21$\pm$.23 (4.00) & .17$\pm$.22 (4.33) \\
& \textsc{DeepArch} & \cellcolor{green!20}1.0$\pm$.00 (1.00) & 1.0$\pm$.00 (2.33) & 1.0$\pm$.00 (2.33) & .56$\pm$.29 (5.00) & .56$\pm$.28 (5.00) & .70$\pm$.18 (3.67) & .55$\pm$.29 (5.00) & .59$\pm$.29 (4.67) & .56$\pm$.29 (5.00) & .57$\pm$.25 (4.67) & .34$\pm$.30 (6.00) \\
& \textsc{Jump3r} & \cellcolor{green!20}.17$\pm$.32 (1.00) & .08$\pm$.22 (1.67) & .08$\pm$.22 (1.67) & \textcolor{red}{\ding{55}} & \textcolor{red}{\ding{55}} & .08$\pm$.22 (1.67) & .08$\pm$.22 (1.67) & .08$\pm$.22 (1.67) & .07$\pm$.20 (2.00) & .03$\pm$.15 (2.67) & .03$\pm$.15 (2.67) \\
& \textsc{HSMGP} & \cellcolor{green!20}.98$\pm$.09 (1.00) & .80$\pm$.27 (2.00) & .72$\pm$.33 (3.00) & .60$\pm$.39 (4.33) & .59$\pm$.40 (4.33) & .68$\pm$.34 (3.33) & .60$\pm$.39 (4.33) & .58$\pm$.40 (4.33) & .66$\pm$.35 (3.33) & .57$\pm$.37 (4.33) & .53$\pm$.38 (4.33) \\

\hline

\multirow{9}{*}{$50\%$}
& \textsc{7z} & \cellcolor{green!20}.86$\pm$.15 (1.00) & .54$\pm$.22 (2.33) & .54$\pm$.23 (2.33) & .56$\pm$.23 (2.33) & .54$\pm$.23 (2.33) & .54$\pm$.23 (2.33) & .54$\pm$.23 (2.33) & .49$\pm$.28 (3.33) & .56$\pm$.24 (2.00) & .36$\pm$.11 (4.67) & .41$\pm$.15 (3.67) \\
& \textsc{Kanzi} & \cellcolor{green!20}.59$\pm$.17 (1.00) & .52$\pm$.16 (2.00) & .52$\pm$.16 (2.00) & \textcolor{red}{\ding{55}} & \textcolor{red}{\ding{55}} & .52$\pm$.16 (2.00) & .51$\pm$.16 (2.33) & .52$\pm$.16 (2.00) & .52$\pm$.16 (2.00) & .37$\pm$.18 (3.33) & .37$\pm$.18 (3.33) \\
& \textsc{ExaStencils} & \cellcolor{green!20}.95$\pm$.07 (1.00) & .54$\pm$.14 (2.67) & .58$\pm$.11 (2.00) & .29$\pm$.16 (4.67) & .29$\pm$.16 (4.67) & .29$\pm$.16 (4.67) & .28$\pm$.17 (4.67) & .37$\pm$.16 (3.67) & .34$\pm$.18 (3.67) & .12$\pm$.07 (5.67) & .11$\pm$.09 (6.33) \\
& \textsc{Apache} & .58$\pm$.09 (2.67) & .59$\pm$.19 (2.33) & .59$\pm$.15 (2.67) & .40$\pm$.15 (4.67) & .39$\pm$.15 (4.67) & .42$\pm$.16 (4.33) & .41$\pm$.18 (4.67) & \cellcolor{green!20}.69$\pm$.17 (1.00) & .40$\pm$.15 (4.67) & \cellcolor{green!20}.72$\pm$.17 (1.00) & \cellcolor{green!20}.69$\pm$.17 (1.00) \\
& \textsc{SQLite} & \cellcolor{green!20}.74$\pm$.19 (1.00) & .53$\pm$.17 (2.33) & .52$\pm$.17 (2.33) & \textcolor{red}{\ding{55}} & \textcolor{red}{\ding{55}} & .52$\pm$.17 (2.33) & .53$\pm$.17 (2.33) & .52$\pm$.16 (2.33) & .54$\pm$.17 (2.00) & .27$\pm$.14 (3.33) & .27$\pm$.14 (3.33) \\
& \textsc{DConvert} & \cellcolor{green!20}.92$\pm$.03 (1.00) & .52$\pm$.28 (2.00) & .41$\pm$.24 (3.00) & .43$\pm$.23 (3.00) & .41$\pm$.24 (3.00) & .43$\pm$.23 (3.00) & .41$\pm$.24 (3.00) & .44$\pm$.22 (2.67) & .41$\pm$.23 (3.00) & .31$\pm$.20 (4.00) & .29$\pm$.20 (4.33) \\
& \textsc{DeepArch} & \cellcolor{green!20}1.0$\pm$.00 (1.00) & 1.0$\pm$.00 (2.33) & 1.0$\pm$.00 (2.33) & .84$\pm$.10 (3.67) & .83$\pm$.12 (3.67) & .84$\pm$.10 (3.67) & .83$\pm$.11 (3.67) & .83$\pm$.12 (3.67) & .83$\pm$.12 (3.67) & .82$\pm$.06 (4.00) & .76$\pm$.15 (5.00) \\
& \textsc{Jump3r} & \cellcolor{green!20}.41$\pm$.27 (1.00) & .27$\pm$.18 (2.00) & .28$\pm$.18 (2.00) & \textcolor{red}{\ding{55}} & \textcolor{red}{\ding{55}} & .28$\pm$.18 (2.00) & .28$\pm$.18 (2.00) & .28$\pm$.18 (2.00) & .27$\pm$.18 (2.00) & .18$\pm$.14 (3.00) & .18$\pm$.14 (3.00) \\
& \textsc{HSMGP} & \cellcolor{green!20}.99$\pm$.04 (1.00) & .84$\pm$.17 (2.00) & .81$\pm$.19 (2.33) & .74$\pm$.23 (3.67) & .73$\pm$.24 (4.00) & .82$\pm$.16 (2.33) & .74$\pm$.23 (4.00) & .76$\pm$.22 (3.67) & .75$\pm$.22 (3.33) & .71$\pm$.22 (4.67) & .70$\pm$.22 (4.67) \\

      \hline

\multirow{9}{*}{$90\%$}
& \textsc{7z} & \cellcolor{green!20}.84$\pm$.24 (1.00) & .44$\pm$.27 (2.00) & .44$\pm$.27 (2.00) & .44$\pm$.27 (2.00) & .44$\pm$.27 (2.00) & .44$\pm$.27 (2.00) & .44$\pm$.27 (2.00) & .41$\pm$.30 (2.33) & .44$\pm$.27 (2.00) & .23$\pm$.10 (4.33) & .28$\pm$.14 (3.33) \\
& \textsc{Kanzi} & \cellcolor{green!20}.35$\pm$.23 (1.00) & .30$\pm$.19 (1.67) & .29$\pm$.19 (1.67) & \textcolor{red}{\ding{55}} & \textcolor{red}{\ding{55}} & .29$\pm$.19 (1.67) & .29$\pm$.19 (1.67) & .29$\pm$.19 (1.67) & .29$\pm$.19 (1.67) & .15$\pm$.14 (2.67) & .15$\pm$.14 (2.67) \\
& \textsc{ExaStencils} & \cellcolor{green!20}.96$\pm$.07 (1.00) & .58$\pm$.16 (2.33) & .58$\pm$.13 (2.33) & .35$\pm$.14 (4.67) & .35$\pm$.13 (4.67) & .34$\pm$.15 (4.67) & .34$\pm$.15 (4.67) & .39$\pm$.12 (3.67) & .37$\pm$.15 (4.00) & .20$\pm$.11 (5.67) & .18$\pm$.07 (6.00) \\
& \textsc{Apache} & .92$\pm$.00 (3.33) & .93$\pm$.02 (3.33) & \cellcolor{green!20}.95$\pm$.04 (1.00) & .92$\pm$.01 (4.33) & .92$\pm$.01 (4.33) & .92$\pm$.01 (4.33) & .92$\pm$.02 (4.00) & .92$\pm$.01 (4.33) & .92$\pm$.01 (4.33) & .94$\pm$.03 (2.00) & .94$\pm$.03 (2.00) \\
& \textsc{SQLite} & \cellcolor{green!20}.70$\pm$.19 (1.00) & .52$\pm$.16 (2.00) & .52$\pm$.16 (2.00) & \textcolor{red}{\ding{55}} & \textcolor{red}{\ding{55}} & .53$\pm$.16 (2.00) & .53$\pm$.16 (2.00) & .54$\pm$.16 (2.00) & .53$\pm$.16 (2.00) & .32$\pm$.08 (3.00) & .32$\pm$.08 (3.00) \\
& \textsc{DConvert} & \cellcolor{green!20}.90$\pm$.07 (1.00) & .55$\pm$.29 (2.00) & .52$\pm$.27 (2.33) & .52$\pm$.27 (2.33) & .52$\pm$.27 (2.33) & .52$\pm$.27 (2.33) & .52$\pm$.27 (2.33) & .53$\pm$.27 (2.33) & .53$\pm$.27 (2.33) & .36$\pm$.25 (3.33) & .36$\pm$.25 (3.33) \\
& \textsc{DeepArch} & \cellcolor{green!20}1.0$\pm$.00 (1.00) & 1.0$\pm$.00 (2.00) & 1.0$\pm$.00 (2.33) & .83$\pm$.12 (3.33) & .83$\pm$.12 (3.33) & .83$\pm$.12 (3.33) & .83$\pm$.12 (3.33) & .84$\pm$.11 (3.33) & .83$\pm$.12 (3.33) & .80$\pm$.10 (4.33) & .76$\pm$.15 (5.00) \\
& \textsc{Jump3r} & \cellcolor{green!20}.23$\pm$.25 (1.00) & .14$\pm$.19 (1.67) & .15$\pm$.21 (1.67) & \textcolor{red}{\ding{55}} & \textcolor{red}{\ding{55}} & .15$\pm$.21 (1.67) & .15$\pm$.21 (1.67) & .15$\pm$.21 (1.67) & .15$\pm$.21 (1.67) & .09$\pm$.15 (2.67) & .09$\pm$.15 (2.67) \\
& \textsc{HSMGP} & \cellcolor{green!20}.99$\pm$.04 (1.00) & .81$\pm$.24 (2.00) & .77$\pm$.27 (2.33) & .66$\pm$.34 (4.00) & .65$\pm$.35 (4.33) & .66$\pm$.34 (4.00) & .65$\pm$.34 (4.33) & .68$\pm$.33 (3.67) & .72$\pm$.30 (3.33) & .59$\pm$.34 (5.00) & .58$\pm$.34 (5.00) \\

      \hline
\multicolumn{2}{c}{Average $p_t$ score/rank}
&\multicolumn{1}{l}{\cellcolor{green!20}.58/1.23}&\multicolumn{1}{l}{.36/2.09}&\multicolumn{1}{l}{.34/2.22}&\multicolumn{1}{l}{.29/3.57}&\multicolumn{1}{l}{.28/3.61}&\multicolumn{1}{l}{.30/2.56}&\multicolumn{1}{l}{.26/2.99} &\multicolumn{1}{l}{.29/2.59}&\multicolumn{1}{l}{.26/2.87}&\multicolumn{1}{l}{.21/3.27}&\multicolumn{1}{l}{.20/3.48} \\

      \bottomrule
    \end{tabular}
  \end{adjustbox}
\vspace{-0.3cm}
\end{table*}

\subsubsection{Results}


\revision{From Table~\ref{tab:rq2}, we clearly observe the superiority of \textsc{CoTune} over the others with/without using requirement as the tuning objective: out of the 54, it has $89\%$ best-ranked cases ($87\%$ sole best) compared with the  $2\%$/$2\%$ ($2\%$/$2\%$ sole best) for \texttt{HEBO}$_p$/\texttt{HEBO}$_r$;  none as best for \texttt{Flash}$_p$/\texttt{Flash}$_r$ (\texttt{Flash} has failed to terminate reasonably for some systems with large space due to its exhaustive nature); none as best for \texttt{SMAC}$_p$/\texttt{SMAC}$_r$; $6\%$/$0\%$ ($4\%$/$0\%$ sole best) for \texttt{TurBO}$_p$/\texttt{TurBO}$_r$; and $4\%$/$4\%$ ($4\%$/$4\%$ sole best) for \texttt{Bounce}$_p$/\texttt{Bounce}$_r$. \textsc{CoTune} also has the best average satisfaction and ranks with an improvement up to $2.9\times$ on satisfaction (against \texttt{Bounce}$_p$). For existing tuners, tuning with and without a requirement do not differ much. This is because of their model-based nature: the uncertain model prediction has weakened the positive impact that could have been brought by the requirement as well as the harm raised from the loss of convergence and stagnating at local optima.}

From Figure~\ref{fig:rq2}, we see that \tuner~needs some limited resources to correctly co-evolve the auxiliary proposition initially ($\approx20$ budget), but it then performs constantly better than the others over different budgets, which confirms its efficiency.

\begin{rbox}
   \textit{\textbf{RQ2:} \tuner~considerably outperforms the state-of-the art tuners: it is ranked first for $89\%$ cases ($87\%$ sole first) against the $2\%$ cases of the overall second best \texttt{HEBO}, with up to $2.9\times$ improved satisfaction and higher efficiency.}
\end{rbox}

\subsection{Ablation Study}

\subsubsection{Method}

To answer \textbf{RQ3}, we assess several variants of \tuner: \tuner$_0$, \tuner$_1$, and \tuner$_2$ that only evolves $p_a$ under \textbf{Case 0}, \textbf{Case 1}, and \textbf{Case 2}, respectively. We compare them with \texttt{GA}$_p$, i.e., \tuner~without any co-evolution. All other settings are the same as \textbf{RQ1} and \textbf{RQ2}.

\subsubsection{Results}

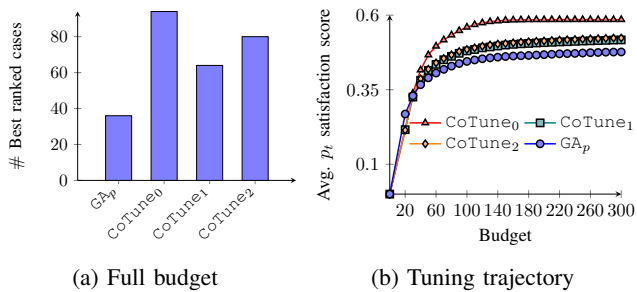
\begin{figure}[t!]
\centering
\subfloat[Full budget]{
\begin{tikzpicture}
  \begin{axis}[
    ybar,                             
    bar width=15pt,   
       axis x line  = bottom,
            axis y line  = left  ,
            y axis line style={black},
            yticklabel style={black},
             width=6cm, height=5cm,
    enlarge x limits=0.3,             
 ylabel={{$\#$ Best ranked cases}},                 
      symbolic x coords={\texttt{GA}$_p$,\texttt{CoTune}$_0$,\texttt{CoTune}$_1$,\texttt{CoTune}$_2$},
    xtick=data,                       
    xticklabel style={rotate=45,anchor=east},
    ymin=0,                           
    legend style={at={(0.5,1.05)}, draw=none,anchor=south, legend columns=2},
    legend cell align={left}
  ]

  \addplot+[style={fill=blue!50,draw=black},error bars/.cd, 
y dir=both,y explicit]
coordinates {
     (\texttt{GA}$_p$,36)
    (\texttt{CoTune}$_0$,94)
    (\texttt{CoTune}$_1$,64)
    (\texttt{CoTune}$_2$,80)};

  \end{axis}

\end{tikzpicture}
}
~\hspace{-0.3cm}
\subfloat[Tuning trajectory]{
\begin{tikzpicture}
    \begin{axis}[
      axis x line  = bottom,
            axis y line  = left  ,
        width=6cm, height=5cm, 
          xlabel={Budget},
         ylabel={Avg. ${p}_t$ satisfaction score},
        xmin=0, xmax=300,
        xtick={20,60,100,140,180,220,260,300},
        ymin=0, ymax=0.6,
        ytick={0.1, 0.35, 0.6},
        grid=minor,
        legend style={
          draw=none, fill=none,
          at={(1.1,0.2)}, anchor=south east,
          legend cell align=left, legend columns=2
        }
    ]


\addplot[
    red,
    mark=triangle*,mark options={fill=red!50,draw=black},
    thick
] coordinates {
(0,0)
(20,0.21524696)
(30,0.33815211)
(40,0.41612114)
(50,0.46677452)
(60,0.49607852)
(70,0.51689166)
(80,0.53345646)
(90,0.54805966)
(100,0.55962126)
(110,0.56817950)
(120,0.57426655)
(130,0.57833939)
(140,0.58068046)
(150,0.58272648)
(160,0.58357639)
(170,0.58412580)
(180,0.58468120)
(190,0.58481825)
(200,0.58488373)
(210,0.58495948)
(220,0.58501716)
(230,0.58502467)
(240,0.58507765)
(250,0.58507765)
(260,0.58507765)
(270,0.58507765)
(280,0.58507765)
(290,0.58507765)
(300,0.58507765)
};\addlegendentry{\texttt{CoTune}$_0$}

\addplot[
    teal,
     mark=square*,mark options={fill=teal!50,draw=black},
    thick
] coordinates {
(0,0)
(20,0.21592082)
(30,0.32528645)
(40,0.38172628)
(50,0.41461054)
(60,0.43487847)
(70,0.45045227)
(80,0.46195385)
(90,0.47208466)
(100,0.48012568)
(110,0.48488368)
(120,0.49029372)
(130,0.49347871)
(140,0.49662598)
(150,0.49974694)
(160,0.50209459)
(170,0.50388402)
(180,0.50510339)
(190,0.50672124)
(200,0.50776234)
(210,0.50902437)
(220,0.51018243)
(230,0.51095100)
(240,0.51216912)
(250,0.51317280)
(260,0.51394347)
(270,0.51475885)
(280,0.51555225)
(290,0.51647159)
(300,0.51703389)
};\addlegendentry{\texttt{CoTune}$_1$}

   \addplot[
    orange,
    mark=diamond*,mark options={fill=orange!50,draw=black},
    thick
] coordinates {
(0,0)
(20,0.21429319)
(30,0.33319383)
(40,0.38662851)
(50,0.41887172)
(60,0.44010030)
(70,0.45465696)
(80,0.46692267)
(90,0.47795908)
(100,0.48476291)
(110,0.49055947)
(120,0.49586028)
(130,0.50014116)
(140,0.50297387)
(150,0.50488873)
(160,0.50727359)
(170,0.50940825)
(180,0.51090520)
(190,0.51224451)
(200,0.51351983)
(210,0.51510873)
(220,0.51636614)
(230,0.51731391)
(240,0.51844197)
(250,0.51945560)
(260,0.52050305)
(270,0.52124368)
(280,0.52226201)
(290,0.52305570)
(300,0.52330836)
};\addlegendentry{\texttt{CoTune}$_2$}

   \addplot[
    blue,
    mark=*,mark options={fill=blue!50,draw=black},
    thick
] coordinates {
(0,0)
(20,0.26853721)
(30,0.32950946)
(40,0.36682536)
(50,0.38968223)
(60,0.40593088)
(70,0.41837047)
(80,0.42917769)
(90,0.43781918)
(100,0.44495712)
(110,0.44976665)
(120,0.45376652)
(130,0.45676914)
(140,0.45926741)
(150,0.46131584)
(160,0.46290411)
(170,0.46394326)
(180,0.46490170)
(190,0.46628323)
(200,0.46755899)
(210,0.46885283)
(220,0.47035093)
(230,0.47113508)
(240,0.47174835)
(250,0.47303315)
(260,0.47420703)
(270,0.47488600)
(280,0.47588090)
(290,0.47619247)
(300,0.47657766)
};\addlegendentry{\texttt{GA}$_p$}

    \end{axis}
\end{tikzpicture}
}

\caption{Ablation analysis of \Model~over all cases/runs.}
\label{fig:rq3}
\vspace{-0.3cm}
\end{figure}

As shown in Figure~\ref{fig:rq3}a, all variants of \tuner~performs generally better than \texttt{GA}$_p$: they have $94$ (\tuner$_0$), $64$ (\tuner$_1$), and $80$ (\tuner$_2$) best ranked cases against the $36$ for \texttt{GA}$_p$---up to $2.61\times$ better. The results also imply that \textbf{Case 0} and \textbf{Case 2} are more devastating than \textbf{Case 1}. For the trajectory in Figure~\ref{fig:rq3}b, all variants need some resources to adapt to the right direction at 20 budget, after which they perform constantly better than the one without co-evolution---a more promising efficiency.

\begin{rbox}
   \textit{\textbf{RQ3:} The co-evolution strategies that handles all three cases in \tuner~are beneficial while handling \textbf{Case 0} and \textbf{Case 2} are more impactful than that of \textbf{Case 1}.}
\end{rbox}

\subsection{RQ4: Sensitivity of \Model~to k}


\subsubsection{Method}

The key parameter for \tuner~is $k$, which sets how many iterations without improvement we need to observe to confirm stagnation. We examine \tuner~under different $k$ values---$k \in \{3,5,7,10\}$---with the same settings as before.


\begin{figure}[t!]
\centering
\subfloat[$d \in \{0.1\%,1\%,5\%\}$]{
\begin{tikzpicture}
  \begin{axis}[
    ybar,                             
    bar width=15pt,   
       axis x line  = bottom,
            axis y line  = left  ,
            y axis line style={black},
            yticklabel style={black},
             width=6cm, height=5cm,
    enlarge x limits=0.3,             
 ylabel={{$\#$ Best ranked cases}},                 
    symbolic x coords={$k=3$,$k=5$,$k=7$,$k=10$},
    xtick=data,                       
    xticklabel style={rotate=45,anchor=east},
    ymin=0,                           
    legend style={at={(0.5,1.05)}, draw=none,anchor=south, legend columns=2},
    legend cell align={left}
  ]

  \addplot+[style={fill=blue!50,draw=black},error bars/.cd, 
y dir=both,y explicit]
coordinates {
    ($k=3$,41) +- (0.0, 0.0)
    ($k=5$,37) +- (0.0, 0.0)
    ($k=7$,42) +- (0.0, 0.0)
    ($k=10$,40) +- (0.0, 0.0)};

  \end{axis}

\end{tikzpicture}
}
~\hfill
\subfloat[$d \in \{20\%,50\%,90\%\}$]{
\begin{tikzpicture}
  \begin{axis}[
    ybar,                             
    bar width=15pt,   
       axis x line  = bottom,
            axis y line  = left  ,
            y axis line style={black},
            yticklabel style={black},
             width=6cm, height=5cm,
    enlarge x limits=0.3,             
 ylabel={{$\#$ Best ranked cases}},                 
    symbolic x coords={$k=3$,$k=5$,$k=7$,$k=10$},
    xtick=data,                       
    xticklabel style={rotate=45,anchor=east},
    ymin=0,                           
    legend style={at={(0.5,1.05)}, draw=none,anchor=south, legend columns=2},
    legend cell align={left}
  ]

  \addplot+[style={fill=blue!50,draw=black},error bars/.cd, 
y dir=both,y explicit]
coordinates {
    ($k=3$,44) +- (0.0, 0.0)
    ($k=5$,38) +- (0.0, 0.0)
    ($k=7$,33) +- (0.0, 0.0)
    ($k=10$,35) +- (0.0, 0.0)};

  \end{axis}

\end{tikzpicture}
}

\caption{Sensitivity of \tuner~to stagnation indicator $k$.}
\label{fig:rq4}
\vspace{-0.3cm}
\end{figure}
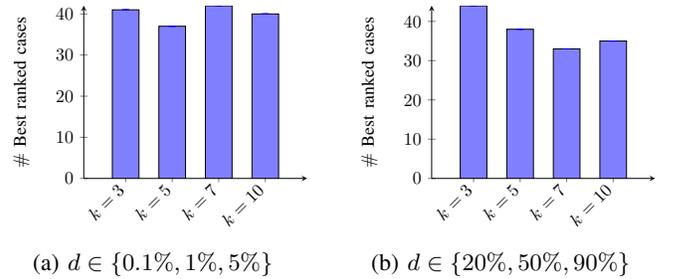

\subsubsection{Results}

In Figure~\ref{fig:rq4}, we see that $k=3$ is generally better than the others due to its more excessive stagnation mitigation. Yet, for harder requirements with $d \in \{0.1\%,1\%,5\%\}$ (Figure~\ref{fig:rq4}a), the benefit of $k=3$ become more blurred against the results of easier requirements (Figure~\ref{fig:rq4}b). This is because in those cases the stagnation in \textbf{Case 2} is much less likely to occur, since most commonly \textbf{Case 0} and \textbf{Case 1} are the main issues caused by the target performance requirement.


\begin{rbox}
   \textit{\textbf{RQ4:} $k=3$ is generally better for more excessive stagnation mitigation, especially on hard requirements.}
\end{rbox}

\section{Discussion}
\label{sec:discussion}

\subsection{Behind the Scenes: Why \tuner~Works?}

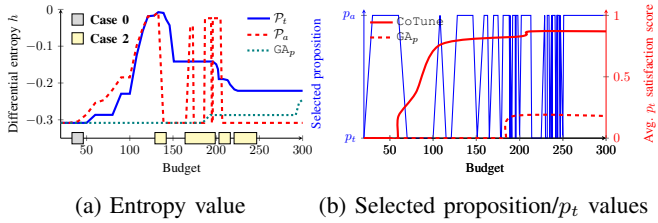
\begin{figure}[!t]
 \centering
  \begin{subfigure}[t]{0.47\columnwidth}
    \centering
\begin{tikzpicture}
    \begin{axis}[
      axis x line  = bottom,
            axis y line  = left  ,
        width=8cm, height=5cm, 
        xlabel={Budget},
        ylabel={Differential entropy $h$},
        xmin=20, xmax=300,
        ymin=-0.35, ymax=0,
        ytick={ -0.30,-0.20,-0.10,0},
            legend cell align=left,
          legend columns=1,
        legend style={
          draw=none, fill=none,
          at={(1,0.6)}, anchor=south east,
          legend cell align=left
        }
    ]

\addplot[blue,
        mark=none,
        ultra thick] coordinates {(20,-0.3084)
(30,-0.3084) (40,-0.3084) (50,-0.3084) (60,-0.2866) (70,-0.2866) (80,-0.2866) (90,-0.2292) (100,-0.2292) (108,-0.1244) (115,-0.0588) (121,-0.0186) (128,-0.0168) (133,-0.0074) (139,-0.0095) (142,-0.0269) (146,-0.0430) (151,-0.1416) (159,-0.1416) (161,-0.1416) (163,-0.1416) (165,-0.1416) (170,-0.1416) (171,-0.1416) (173,-0.1416) (173,-0.1416) (174,-0.1416) (175,-0.1416) (178,-0.1416) (179,-0.1416) (181,-0.1416) (183,-0.1416) (183,-0.1416) (186,-0.1416) (187,-0.1416) (188,-0.1416) (188,-0.1416) (189,-0.1416) (189,-0.1416) (189,-0.1416) (191,-0.1416) (192,-0.1416) (192,-0.1416) (193,-0.1416) (194,-0.1416) (194,-0.1416) (195,-0.1416) (195,-0.1416) (195,-0.1416) (196,-0.1416) (196,-0.1416) (198,-0.1453) (200,-0.1492) (202,-0.1623) (203,-0.1893) (205,-0.1893) (207,-0.1893) (208,-0.1893) (213,-0.1921) (214,-0.1949) (216,-0.2040) (221,-0.2175) (225,-0.2212) (229,-0.2212) (229,-0.2212) (230,-0.2212) (232,-0.2212) (232,-0.2212) (233,-0.2212) (234,-0.2212) (234,-0.2212) (234,-0.2212) (234,-0.2212) (234,-0.2212) (235,-0.2212) (237,-0.2212) (237,-0.2212) (237,-0.2212) (238,-0.2212) (238,-0.2212) (241,-0.2212) (242,-0.2212) (243,-0.2212) (244,-0.2212) (244,-0.2212) (244,-0.2212) (244,-0.2212) (245,-0.2212) (246,-0.2212) (246,-0.2212) (246,-0.2212) (246,-0.2212) (246,-0.2212) (247,-0.2212) (247,-0.2212) (247,-0.2212) (247,-0.2212) (250,-0.2212) (250,-0.2212) (251,-0.2212) (251,-0.2212) (300,-0.2212)
}; 

    \addplot[red,
        mark=none,
        ultra thick,dashed] coordinates {(20,-0.3084)
(30,-0.3084) (40,-0.3043) (50,-0.2853) (60,-0.2542) (70,-0.2463) (80,-0.2185) (90,-0.1847) (100,-0.1846) (108,-0.0980) (115,-0.0508) (121,-0.0190) (128,-0.0171) (133,-0.0079) (139,-0.3084) (142,-0.3084) (146,-0.3084) (151,-0.3084) (159,-0.3084) (161,-0.3084) (163,-0.3084) (165,-0.3084) (170,-0.0457) (171,-0.0457) (173,-0.0457) (173,-0.0457) (174,-0.3084) (175,-0.3084) (178,-0.3084) (179,-0.3084) (181,-0.3084) (183,-0.3084) (183,-0.3084) (186,-0.3084) (187,-0.0239) (188,-0.0239) (188,-0.0239) (189,-0.0239) (189,-0.0239) (189,-0.0239) (191,-0.0239) (192,-0.0239) (192,-0.0239) (193,-0.0239) (194,-0.0239) (194,-0.0239) (195,-0.3084) (195,-0.3084) (195,-0.3084) (196,-0.3084) (196,-0.0239) (198,-0.0239) (200,-0.0239) (202,-0.0239) (203,-0.0239) (205,-0.0239) (207,-0.3084) (208,-0.3084) (213,-0.3084) (214,-0.3084) (216,-0.3084) (221,-0.3084) (225,-0.3084) (229,-0.3084) (229,-0.3084) (230,-0.3084) (232,-0.3084) (232,-0.3084) (233,-0.3084) (234,-0.3084) (234,-0.3084) (234,-0.3084) (234,-0.3084) (234,-0.3084) (235,-0.3084) (237,-0.3084) (237,-0.3084) (237,-0.3084) (238,-0.3084) (238,-0.3084) (241,-0.3084) (242,-0.3084) (243,-0.3084) (244,-0.3084) (244,-0.3084) (244,-0.3084) (244,-0.3084) (245,-0.3084) (246,-0.3084) (246,-0.3084) (246,-0.3084) (246,-0.3084) (246,-0.3084) (247,-0.3084) (247,-0.3084) (247,-0.3084) (247,-0.3084) (250,-0.3084) (250,-0.3084) (251,-0.3084) (251,-0.3084) (300,-0.3084)
};  

 \addplot[teal,
        mark=none,
        ultra thick,dotted] coordinates
{
(20,-0.3084) (30,-0.3084) (40,-0.3084) (50,-0.3084) (59,-0.3084) (68,-0.3084) (76,-0.3084) (85,-0.3084) (92,-0.3084) (101,-0.3084) (108,-0.3084) (116,-0.3084) (126,-0.3084) (135,-0.3084) (142,-0.3084) (151,-0.3084) (158,-0.3084) (166,-0.3084) (171,-0.3084) (178,-0.3084) (184,-0.3084) (193,-0.2871) (202,-0.2871) (210,-0.2871) (216,-0.2871) (224,-0.2871) (232,-0.2871) (241,-0.2871) (250,-0.2871) (258,-0.2871) (265,-0.2871) (271,-0.2871) (278,-0.2871) (287,-0.2871) (292,-0.2871) (296,-0.2527) (304,-0.2387)
}; 

\addlegendentry{$\mathcal{P}_t$}
\addlegendentry{$\mathcal{P}_a$}
\addlegendentry{\texttt{GA}$_p$}
    \end{axis}

    \draw [draw=black,fill=gray!30] (0.6,0.15) rectangle (0.3,-0.15);
    \draw [draw=black,fill=yellow!30] (2.5,0.15) rectangle (2.8,-0.15);
    \draw [draw=black,fill=yellow!30] (3.3,0.15) rectangle (4.1,-0.15);
    \draw [draw=black,fill=yellow!30] (4.2,0.15) rectangle (4.5,-0.15);
    \draw [draw=black,fill=yellow!30] (4.6,0.15) rectangle (5.2,-0.15);

     \draw [draw=black,fill=gray!30] (0.6,3.3) rectangle (0.3,3);
      \draw [draw=black,fill=yellow!30] (0.6,2.8) rectangle (0.3,2.5);

      \node[draw=none] at (1.3,3.15) {\textbf{Case 0}};
      \node[draw=none] at (1.3,2.65) {\textbf{Case 2}};
\end{tikzpicture}
  \subcaption{Entropy value}
  \end{subfigure}
 ~\hspace{-0.5cm}
  \begin{subfigure}[t]{0.53\columnwidth}
    \centering
\begin{tikzpicture}
    \begin{axis}[
      axis x line  = bottom,
            axis y line  = left  ,
        width=8cm, height=5cm, 
        xlabel={Budget},
          y axis line style={blue},
            yticklabel style={blue},
        ylabel={\textcolor{blue}{Selected proposition}},
        xmin=20, xmax=300,
        ymin=0, ymax=1.05,
        ytick={0,1},
          yticklabels={$p_t$,$p_a$},
            legend cell align=left,
          legend columns=1,
        legend style={
          draw=none, fill=none,
          at={(1.0,0.6)}, anchor=south east,
          legend cell align=left
        }
    ]

    \addplot[
        blue,
        mark=none,
     thick
    ] coordinates { (20,0)
(30,1) (40,1) (50,1) (60,1) (70,0) (80,0) (90,0) (100,0) (108,1) (115,0) (121,0) (128,1) (133,1) (139,1) (142,1) (146,1) (151,0) (159,1) (161,1) (163,1) (165,0) (170,1) (171,1) (173,1) (173,1) (174,1) (175,1) (178,0) (179,0) (181,1) (183,1) (183,1) (186,1) (187,1) (188,1) (188,1) (189,0) (189,1) (189,1) (191,0) (192,1) (192,1) (193,1) (194,1) (194,1) (195,1) (195,1) (195,0) (196,0) (196,1) (198,1) (200,0) (202,1) (203,1) (205,1) (207,1) (208,1) (213,1) (214,0) (216,0) (221,0) (225,0) (229,1) (229,1) (230,0) (232,0) (232,1) (233,1) (234,1) (234,1) (234,1) (234,1) (234,1) (235,0) (237,1) (237,1) (237,0) (238,0) (238,1) (241,1) (242,0) (243,1) (244,1) (244,1) (244,0) (244,1) (245,1) (246,1) (246,1) (246,1) (246,1) (246,1) (247,0) (247,1) (247,1) (247,1) (250,0) (250,0) (251,1) (251,1) (300,1)
};

    \end{axis}

     \begin{axis}[
      axis x line  = bottom,
            axis y line  = right  ,
        width=8cm, height=5cm, 
        xlabel={Budget},
          y axis line style={red},
            yticklabel style={red},
             ylabel={\textcolor{red}{Avg. ${p}_t$ satisfaction score}},
        xmin=20, xmax=300,
        ymin=0, ymax=1.05,
            legend cell align=left,
          legend columns=1,
        legend style={
          draw=none, fill=none,
          at={(0.36,0.7)}, anchor=south east,
          legend cell align=left
        }
    ]

    \addplot[
        red,
        mark=none,smooth,
     ultra thick
    ] coordinates { 
(20,0) (30,0) (40,0) (50,0) (59,0) (60,0.18) (80,0.37) (108,0.76) (202,0.83) (213,0.87) (300,0.87)
};

   \addplot[
        red,
        mark=none,smooth,
     ultra thick,dashed
    ] coordinates { 
(20,0) (30,0) (40,0) (50,0) (59,0) (68,0) (76,0) (85,0) (92,0) (101,0) (108,0) (116,0) (126,0) (135,0) (142,0) (151,0) (158,0) (166,0) (171,0) (178,0) (184,0) (193,0.18) (300,0.18)

}; 

\addlegendentry{\texttt{CoTune}}
\addlegendentry{\texttt{GA}$_p$}

    \end{axis}
\end{tikzpicture}
    \subcaption{Selected proposition/$p_t$ values}
  \end{subfigure}
   \caption{A randomly chosen example run for \textsc{SQLite}.}
   \label{fig:dis}
   \vspace{-0.3cm}
 \end{figure}

As from Figure~\ref{fig:dis}a, \tuner~initially encounters \textbf{Case 0} and hence increases the entropy by co-evolving $p_a$; from 145 budget onward it faces several occasions of \textbf{Case 2}, and hence it seeks to keep the entropy minimized via the co-evolution. This influences the evolution of both $\mathcal{P}_a$ and $\mathcal{P}_b$. Therefore, in Figure~\ref{fig:dis}b, the proposition used to guide the tuning is also more frequently switched at the later half than the first half budget. The co-evolution becomes stable from around 250 budgets onward. Without the co-evolution, the \texttt{GA}$_p$ (Figure~\ref{fig:dis}a) exhibits low entropy throughout since most configurations have $p_t=0$ for a long period, and this is only changed at 193 budget point, after which it suffers stagnation. As a result, \texttt{GA}$_p$ achieves only $p_t=0.18$ against the $p_t=0.87$ for \tuner, which also has a much steeper/better trajectory.


\subsection{Threats to Validity}
\label{sec:threats}

To mitigate threats to \textbf{internal validity}, the parameters are pragmatically set based on their implication (e.g., $k$ in \tuner) or follow prior works~\cite{DBLP:journals/tse/Nair0MSA20,chen2025accuracy,DBLP:journals/corr/abs-2112-07303} (e.g., mutation rate and budget), but indeed these can be consolidated.



Since our goal is to satisfy the requirement, we use the systematically quantified satisfaction score as the main metric to ensure \textbf{construct validity}. We have also examined the trajectories under different budget settings with Scott-Knott ESD test. However, unintentional ignorance might still exist.

To ensure \textbf{external validity}, we evaluate \tuner~against four state-of-the-art tuners (with and without requirements) over nine systems from diverse domains and 18 different target performance requirements, leading to 162 cases. Yet, we agree that more subjects might improve the generalizability.

\section{Related Work}
\label{sec:related_work}

\textbf{General Configuration Tuning:} Tuning configurations have been generally done assuming that the better performance would always be more preferred. Genetic Algorithm has been widely used as the core of a tuner~\cite{k2vtune,DBLP:journals/ase/GerasimouCT18,DBLP:journals/corr/abs-2112-07303,DBLP:conf/wosp/MartensKBR10}. Recently, \texttt{MMO}~\cite{DBLP:journals/corr/abs-2112-07303,DBLP:conf/sigsoft/0001L21} is an optimization model that multi-objectvizes configuration tuning, overcoming local optima for the target performance objective via an extra performance metric. Other tuners leverage a surrogate model~\cite{DBLP:conf/icse/XiangChen26,gong2024dividable,DBLP:conf/sigsoft/Gong023,DBLP:journals/pacmse/Gong024,DBLP:journals/tse/ChenB17} to expedite the tuning via a variant of Bayesian optimization~\cite{DBLP:journals/tse/Nair0MSA20,SMAC,DBLP:conf/mascots/JamshidiC16,cowen2022hebo}, e.g., \texttt{PromiseTune}~\cite{DBLP:conf/icse/ChenChen26}, with diverse internal designs. 


All the above tuners have ignored the valuable information in a given performance requirement---the key in \tuner.


\textbf{Performance Requirements Understandings:}
Languages and notations exist for formulating performance requirements~\cite{DBLP:conf/re/EckhardtVFM16,DBLP:journals/re/WhittleSBCB10,DBLP:journals/tosem/ChenL23a}. For example, Eckhardt et al.~\cite{DBLP:conf/re/EckhardtVFM16} present a framework of patterns to systematically interpret the performance requirements. However, they have not systematically quantified them. Chen and Li~\cite{DBLP:journals/tosem/ChenL23a} study the impacts of using performance requirements as an objective for configuration tuning. Their findings suggest that reasonably elicited requirements can help, but poorly specified ones are harmful. Yet, they have neither defined to what extent a performance requirement is appropriate nor proposed any automated tools.

\tuner, on the other hand, is an automated tuner that can robustly leverage the performance requirement. 


\textbf{Requirement-guided Configuration Tuning:} There exist some tuners that incorporate the hard constraint of performance requirement as the objective to guide the configuration tuning using \texttt{GA}~\cite{DBLP:conf/wosp/MartensKBR10,DBLP:journals/ase/GerasimouCT18,DBLP:journals/tsc/ChenB17}. Ghanbari et al.~\cite{DBLP:journals/fgcs/GhanbariSLI12} adopt the same and they assume quadratically quantified requirements. Yet, those tuners differ from \tuner~in the following:

\begin{itemize}
    \item They work on a fixed type of performance requirement while \tuner~has no such a constraint.
    \item They assume a given requirement is always elicited perfectly/reasonably, which is unrealistic. \tuner~works well even if the given target one is harmful to the tuning. 
\end{itemize}

A recent tool~\cite{10.1145/3712185} permits interactive configuration tuning, allowing developers to specify preferences between performance metrics at tuning. Although like \tuner, it does gradually adjust the requirements, the process is however relies on human intervention; \tuner~does so automatically. 


\revision{\textbf{Constrained Tuning:} There exist approaches for constrained configuration tuning~\cite{DBLP:conf/ppopp/0179SMZMDL21,DBLP:conf/asplos/HellstenSLLHEKS23} or general constrained optimization~\cite{DBLP:journals/isci/DAngeloP21,DBLP:conf/aistats/ErikssonP21}. However, those differ from \tuner~in the sense that their constraint definitions are mostly on the dependency constraints/extra constraints in the solution/configuration space\footnote{\revision{\tuner~sets a configuration that violates dependency constraint with the worst possible value, and hence it would be naturally eliminated during tuning iterations. This is the simplest and standard method used in prior work~\cite{DBLP:conf/sigsoft/0001L24}.}}, while our problem places constraint on the same single-objective/performance metric to be optimized, which is complementary to the definition in prior works.}

\section{Conclusion}
\label{sec:conclusion}

This paper presents \tuner, an automated tool that co-evolves a given performance requirement and configurations. Instead of purely leveraging the target performance requirement as an objective or completely ignoring it, \tuner~builds an auxiliary performance requirement and co-evolve it with the configurations to mitigate different situations based on the discriminative power of the tuning.  We show that \tuner~can:


\begin{itemize}
    \item achieve significantly superior results than tuning with static target performance requirement and without;
    \item considerably outperform state-of-the-art tuners regardless if they use the requirement as the objective;
    \item doing so with generally better resource efficiency.
\end{itemize}

The future work extending from \tuner~is fruitful, including the multi-objective cases~\cite{DBLP:journals/tse/LiCY22,DBLP:journals/tosem/ChenL23} and incorporating the information of configuration landscape~\cite{DBLP:conf/seams/Chen22} into the co-evolution.


\section*{Acknowledgement}
This work was supported by a NSFC Grant (62372084) and a UKRI Grant (10054084).

\bibliographystyle{IEEEtran}
\bibliography{references}

\balance

\end{document}